\documentclass[aps,pra,twocolumn]{revtex4}

\usepackage{amsmath, amssymb, amsthm}
\usepackage{amsfonts}
\usepackage{graphicx}
\usepackage{xcolor}
\usepackage[colorlinks=true,
citecolor=blue, 
linkcolor=red]{hyperref}
\usepackage{float}
\newcommand{\C}{\mathcal{C}}
\newcommand{\simp}{\Delta_{\mathcal{X}}}
\newcommand{\A}{\mathcal{A}}
\newcommand{\F}{\mathcal{F}}
\newcommand{\X}{\mathcal{X}}
\newcommand{\Y}{\mathcal{Y}}
\newcommand{\pc}{\mathrm{pc}}
\newcommand{\MMC}{\mathrm{MC}}
\newcommand{\E}{\mathrm{E}}
\newcommand{\Hh}{\mathrm{H}}
\newcommand{\I}{\mathrm{I}}

\newcommand{\MC}{\mathrm{MC}}

\newcommand{\q}{\Tilde{q}}

\newcommand{\x}{\boldsymbol{x}}
\newcommand{\avg}[1]{\langle #1 \rangle}
\newcommand{\Ii}{\Tilde{\mathrm{I}}}

\begin{document}

\title{Minimal Thermodynamic Cost of Communication}

\author{Abhishek Yadav${}^{ab}$}
\author{David Wolpert${}^{acde}$}
\affiliation{\vspace{.25cm}${}^a$Santa Fe Institute, 1399 Hyde Park Road, Santa Fe, NM 87501, USA}
\affiliation{${}^b$Department of Physical Sciences, IISER Kolkata, Mohanpur 741246, India}
\affiliation{${}^c$Complexity Science Hub, Vienna, Austria}
\affiliation{${}^d$Arizona State University, Tempe, Arizona, USA}
\affiliation{${}^e$International Center for Theoretical Physics, Trieste 34151, Italy}

\begin{abstract}

Thermodynamic cost of communication is a major factor in the thermodynamic cost of real-world computers, both biological and digital. Despite its importance, the fundamental principles underlying this cost remain poorly understood. This paper makes two major contributions to addressing this gap. First, we derive a universal relationship between information transmission rate and minimal entropy production (EP) by focusing on the mismatch cost (MMC) component of thermodynamic cost. The resulting relationship holds independently of the underlying physical dynamics, making it broadly applicable. We discuss the implications of the minimal communication cost for work extraction in measurement-and-feedback protocols, and through examples involving binary channels, we show that the relationship between transmission rate and minimal thermodynamic cost can exhibit diminishing returns in certain scenarios.
Second, we extend this thermodynamic analysis to the computational front and back ends critical to  communication---namely, encoding and decoding to reduce errors in noisy transmission. Using the framework of periodic machines, we establish strictly positive minimal costs for implementing linear error-correcting codes. We compare these costs with end-to-end error rates, highlighting trade-offs between thermodynamic efficiency and decoding accuracy.

\end{abstract}

\maketitle
%\tableofcontents
\section{Introduction}

\subsection{Background and Scope}

Modern computers rely on decomposing large-scale, complex problems into simpler calculations handled by sub-components that are reused iteratively throughout a computation.  These components, such as logic gates within circuits, must communicate to execute desired computation. Given that communication is integral to nearly all forms of computation, it becomes imperative to understand the energetic costs associated with communication to fully understand the energetic costs of computation itself. Even if individual computing units are thermodynamically efficient, the overall system can still incur a high thermodynamic cost due to the energy required for communication. 

Early work by Landauer and Bennett investigated the energetic costs of computation~\cite{landauer1961irreversibility, bennett1982thermodynamics, landauer1996physical}, arriving at a semi-formal understanding of minimal heat generation in logically irreversible processes such as bit erasure. These early analyses were based on the idea that a change in the Shannon entropy of a system during computation results in exchange of heat with the environment, quantified by $k_B T \Delta S$, where $k_B$ is the Boltzmann constant, $T$ the temperature of the thermal bath, and $\Delta S$ the change in the system's entropy. While this intuition has been validated by more recent developments in stochastic thermodynamics, it was incomplete~\cite{sagawa2014thermodynamic}.

In particular, it failed to distinguish between reversible and irreversible components of total heat flow. The entropy change $\Delta S$ contributes only to the reversible part of heat transfer—heat that can, in principle, be recovered through a time-reversed protocol~\cite{van2015ensemble}. However, it is now understood that an additional, inherently irreversible contribution exists: the entropy production (EP), which accounts for heat that is unavoidably lost to the environment~\cite{van2015ensemble, seifert2008stochastic}.

This important distinction was not fully appreciated in early work, leading to the inaccurate view that logical reversibility implies thermodynamic reversibility. In reality, these are distinct concepts. A computation is logically reversible if, given any output, the input can be uniquely reconstructed—this condition ensures no loss of information and implies $\Delta S = 0$, yielding zero Landauer cost. However, thermodynamic reversibility requires that the EP be zero. Thus, while logical reversibility eliminates the reversible (Landauer) component of heat, it does not necessarily eliminate the total energetic cost: entropy production may still be present. Logical reversibility alone is therefore insufficient to guarantee thermodynamic reversibility~\cite{sagawa2014thermodynamic}.

This misconception influenced early views on communication as well. In earlier literature, it was argued that communication—viewed as a copy operation—could be achieved without any fundamental energetic cost.~\cite{landauer1996minimal}. For instance, consider the following operation:

\begin{equation}\label{CNOT}
    00 \mapsto 00 \quad \text{and} \quad 10 \mapsto 11.
\end{equation}
Here, the left bit is copied to the right bit, and the right bit always starts in state 0. This operation is logically reversible, and thus was claimed to have no unavoidable thermodynamic cost. But such reasoning only applies to the reversible component of heat flow and ignores EP. 

This paper has two main goals. First, we derive a strictly positive lower bound on the thermodynamic cost associated with communication through a noisy channel, thereby establishing that communication---contrary to some earlier beliefs---has an unavoidable thermodynamic cost. We model communication as a noisy copy operation, analogous to the idealized example in~(\ref{CNOT}), with the channel consisting of an input and an output node. Crucially, the operation of a communication channel involves copying the value of input node to the output node with some noise and subsequently resetting the value of the input node with a new value drawn from a source. This cycle—reset, copy, reset—repeats with each new message.

Second, we extend this thermodynamic analysis to the computational front and back ends critical to  communication i.e., encoding and decoding to reduce errors in noisy transmission~\cite{shannon1948mathematical}. Subsequently, we present a comprehensive analysis of the thermodynamic costs associated with the complete end-to-end communication system, encompassing the encoder, channel, and decoder.

Although prior work has examined energetic costs of encoding and decoding—particularly in the context of VLSI circuits~\cite{grover2011towards, grover2012fundamental}—our approach differs in scope and methodology. We use tools from stochastic thermodynamics and analyze abstract, algorithmic implementations of encoding and decoding, rather than specific hardware realizations. As such, the bounds we present are theoretical, and offer insights into energetic costs of encoding and decoding algorithms. 

Our analysis is grounded in the mismatch cost (MMC) decomposition of thermodynamic cost. For any process on state space $\X$ that evolves an initial distribution $p_0$ to a final distribution $p_1 = Gp_0$, a broad class of thermodynamic cost functions can be written in the form:
\begin{equation}\label{def_cost}
\C(p_0) = \avg{f}_{p_0} + S(Gp_0) - S(p_0),
\end{equation}
where $f: \X \to \mathbb{R}$ is a finite real-valued function and $S(\cdot)$ denotes Shannon entropy.

Such a cost admits the following decomposition~\cite{kolchinsky2021dependence, wolpert2020thermodynamics}:
\begin{equation}\label{def_MMC0}
\C(p_0) = D(p_0 \| q_0) - D(Gp_0 \| Gq_0) + \C(q_0),
\end{equation}
where $D(\cdot \| \cdot)$ is the Kullback–Leibler (KL) divergence, and $q_0 = \arg\min_{r_0} \C(r_0)$ is known as the prior associated with the cost function $\C$. The quantity $D(p_0 \| q_0)~-~D(Gp_0 \| Gq_0)$ is referred to as the mismatch cost (MMC). It quantifies the portion of the total cost attributable to a mismatch between the actual initial distribution $p_0$ and the prior $q_0$. This decomposition holds for any choice of $G$ and $f$, and importantly, MMC provides a non negative contribution for any $p_0 \in \simp$.

Several thermodynamic cost measures---including total entropy production, the change in nonequilibrium free energy, and nonadiabatic entropy production---can be expressed in the form of~(\ref{def_MMC0}) and therefore MMC gives a non-negative contribution in those cases~\cite{kolchinsky2021dependence}. Recent work has shown that the MMC can account for a substantial portion of the total thermodynamic cost in certain scenarios~\cite{yadav2024mismatch}.

In this paper, we specifically focus on the MMC contribution to the total thermodynamic cost. It is important to emphasize that we do not assume or specify any particular physical model for the components of the communication system. Instead, our analysis centers on how the probability distribution over system states evolves during each fundamental step of the communication process. For instance, in the case of encoders and decoders, each computational step induces a change in the system's state distribution, resulting in a sequence $\{p_0, p_1, \ldots, p_n\}$.

It is true that, given a function $f:\X \to \mathbb{R}$ and map $G$, the corresponding prior distribution can be derived~\cite{yadav2024mismatch}. The function $f$ captures some physical details of the process that implements the map $G$. For instance, in the case of total EP, $f(x)$ represents the average heat exchanged between the system and its thermal environment when the system begins in state $x$.

In this work though, we do not derive the prior from a specific physical model of a communication system. Instead, to gain concrete and interpretable insights into the thermodynamics of communication processes, we adopt simple and illustrative choices for the prior distributions.

Nonetheless, the minimal thermodynamic cost that we derive for communication channels holds independently of the specific choice of prior or the physical implementation of the communication channel. The resulting bound is universal and applies to any communication channel.

In the case of encoder and decoder components, however, our chosen priors are not intended to imply generality across all possible priors. Instead, they serve to illustrate that---even in the absence of a detailed physical model---each algorithmic step can contribute a non-zero thermodynamic cost, and MMC can be used to analyze it once the prior is known. 

\subsection{Roadmap and summary}

The paper is organized as follows: we briefly review Shannon's communication theory in Sec.~\ref{sec:info_theory}. We restrict our attention to independent and identically distributed (i.i.d.) information sources, and discrete memoryless communication channels. We present entropy of source as an information measure and mutual information across the channel as the rate of information transfer through a channel. Subsequently, we give a brief account of error-correcting codes, outlining how they enable the detection and correction of message errors, with a specific focus on linear encoders and syndrome decoders.

Sec.~\ref{stoch_thermo} provides an overview of stochastic thermodynamics. We review MMC lower bound on EP, laying the foundation for later discussions. 
In Sec.~\ref{thermo_cost_channels}, we present our first main result: a MMC lower bound on the thermodynamic cost of communication through a noisy channel (Eq.~\ref{eq:MCvsMI} and~\ref{MMC_MI_lower}). We then highlight measurement-and-feedback scenarios where communication cost directly impact work extraction, noting that measurement is essentially a communication process, and as a result, the minimal cost associated with communication plays a crucial role in such settings.

In Sec.~\ref{example_binary_channel}, we examine a concrete example involving binary channels. For a specific choice of prior, we demonstrate that the relationship between the rate of information transmission and the corresponding minimal cost is concave, reflecting diminishing returns. This implies that splitting the information rate across multiple channels can yield a lower total minimal cost than transmitting the same rate through a single channel.

Sec.~\ref{thermo_cost_codes} shifts the focus to the encoding and decoding procedures. These algorithms consist of a sequence of computational steps. In Sec.~\ref{MC_of_Algorithms}, we introduce a formal framework to model the implementation of such algorithms, grounded in the periodic nature of modern computer architectures and the essential role of the instruction pointer. This framework was initially developed in~\cite{yadav2024mismatch}.

Using this framework, Secs.~\ref{sec_MC_linear_codes} and~\ref{sec_MC_syndrome_decoder} analyze the stepwise MMC incurred by linear encoders and syndrome decoders. Importantly, we make some rather straightforward and illustrative choices for the priors, usually assuming that they are uniform distribution. This allows us to compare the minimal thermodynamic cost across various linear error-correcting codes, each characterized by different encoding lengths and decoding error rates. Finally, we combine the thermodynamic cost of encoding and decoding with the minimal cost associated with the communication channel, yielding a total end-to-end cost of the full communication system---comprising encoding, channel transmission, and decoding. We compare the associated MMC cost of the system with the corresponding end-to-end error rate, and highlight the trade-offs between decoding error rates and the minimal cost. We conclude with a broader discussion in Sec.~\ref{Discussion}.

\section{Preliminaries}\label{Preliminaries}

\subsection{Shannon's theory of communication}
\label{sec:info_theory}

\begin{figure}
    \centering
    \includegraphics[trim = {0 3cm 0 0}, width=1\linewidth]{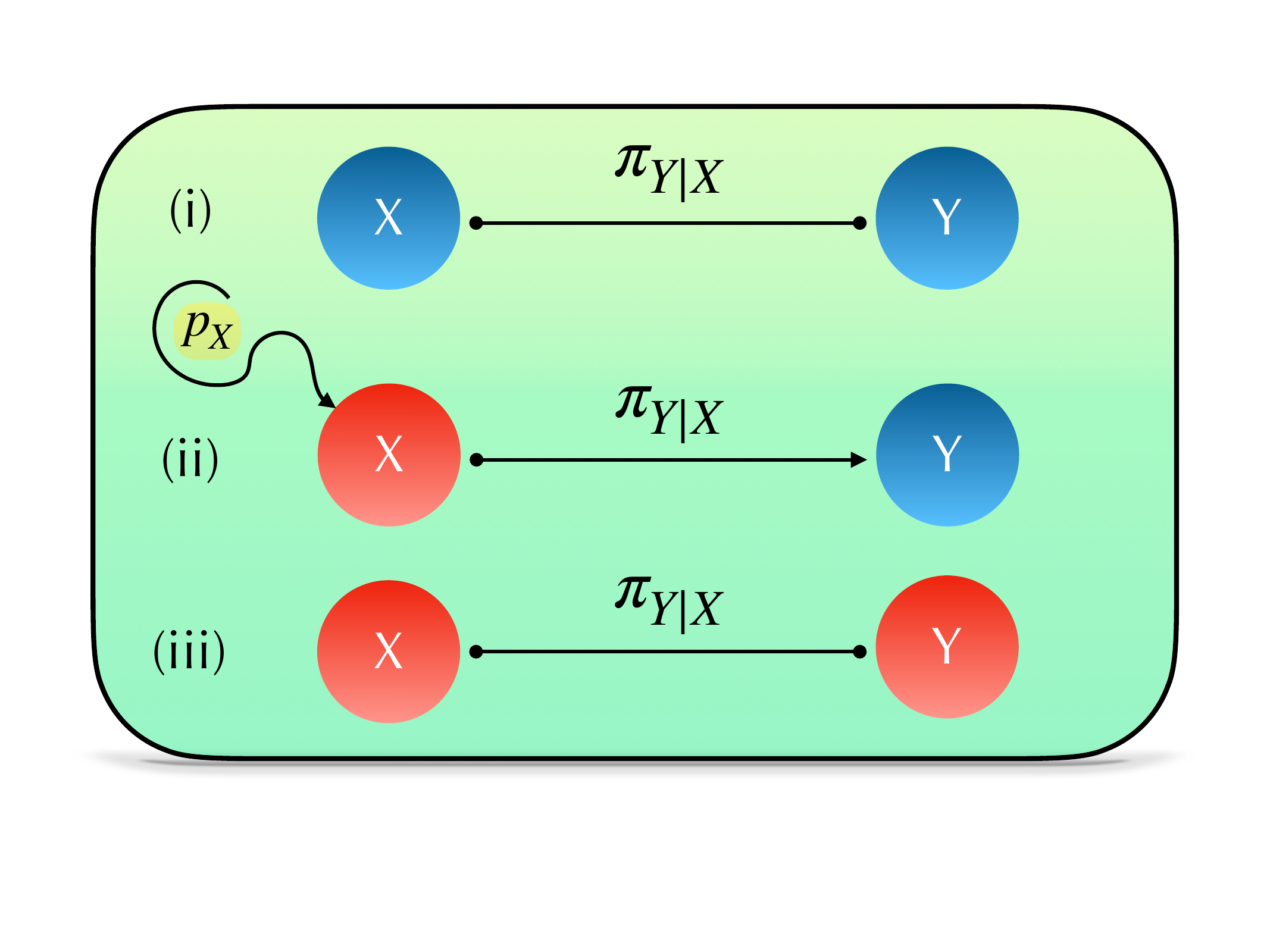}
    \caption{Schematic of a communication channel: (i) The state of the input node $X$ is updated with a new value drawn from the information source. (ii) This input state is then copied to the output node $Y$, subject to noise characterized by the conditional distribution $\pi_{Y|X}$.}
    \label{fig:enter-label}
\end{figure}

The fundamental problem of communication involves transmitting information from a source to a receiver, a process that is almost always subject to unavoidable noise. At an abstract level, communication can be viewed as a copying operation, where symbols generated by a source are reproduced, often imperfectly, at the receiving end.
%{\color{red}\bf Examples: RNA polymerase ...}

Claude Shannon laid the foundation of modern information theory by quantifying the information content of a symbol emitted by a source, or more precisely, the average amount of information generated per symbol~\cite{shannon1948mathematical}. For an i.i.d. source defined by a set of possible messages and a corresponding probability distribution, the average information content per symbol is  quantified by the Shannon entropy:
\begin{equation}
    S(p_X) = - \sum_{x \in \X} p_X(x) \log p_X(x)
\end{equation}
where $\X$ is the set of possible messages, $X$ is the random variable associated with the input state, and $p_X$ is the distribution over the messages. The source is maximally informative when $p_X$ is uniformly distributed over $X$. Conversely, the source produces zero information when the probability distribution is degenerate (i.e., when all probability mass is concentrated on a single symbol). 

If the channel, or the copy operation, is noisy, then the information received is not identical to the information produced at the source; some of it is inevitably lost during transmission. Let the noise in the channel be characterized by the conditional distribution $\pi_{Y|X}$, where $Y$ is the output variable with state space $\Y$. On average, the amount of information lost due to noise is given by the conditional entropy:

\begin{equation}
    S(X|Y) = -\sum_{\substack{x\in \X,\\ y \in \Y}} p_{XY}(x, y) \log \pi_{X|Y}(x|y),
\end{equation}
where $p_{XY}$ is the joint distribution obtained by $p_{XY}(x,y) = \pi_{Y|X}(y) p_X(x)$. The amount of information successfully transmitted from source to receiver is then given by the difference between the total information generated at the source and the information lost in the channel:
\begin{equation}
    I(X; Y) = S(X) - S(X|Y),
\end{equation}
known as the mutual information between $X$ and $Y$.

This mutual information represents the average information transmitted per use of the channel, and for that reason it is called the rate of information transmission or rate of communication. 
%{\color{red}\bf Connect it to other things}.

\subsubsection*{Error Correcting Codes}\label{error_correcting_codes}

In modern communication protocols, the focus is on accurately predicting the original message despite the errors introduced by the channel noise. For this purpose, a communication  channel is often supplemented with an encoder to encode the messages before sending them through the channel and a decoder at the other end of channel~\cite{mackay2003information}. 
Assuming that the source messages are optimally represented as binary strings of length $k$, an $(n, k)$ binary encoder is a function that maps these messages to binary strings of length $n>k$:

\begin{equation}
    \E : \{0, 1\}^k \longrightarrow \{0, 1\}^n
\end{equation}
Each binary string $\mathbf{s} \in \{0, 1\}^k$ is mapped to a \textit{valid codeword} $\E(\mathbf{s})$ in $\{0, 1\}^n$. \textit{Hamming distance} between two codewords is defined as the number of positions at which their bits differ. The \textit{minimum distance} $d_{\min}$ of a code $\E$ is the smallest Hamming distance between any pair of distinct valid codewords:

\begin{equation}
    d_{\min} := \min_{\{\mathbf{s}\ne \mathbf{s}'\}} d(\E(\mathbf{s}), \E(\mathbf{s}'))
\end{equation}
where $d(\mathbf{t}, \mathbf{t}')$ is the Hamming distance between $\mathbf{t}$ and $\mathbf{t}'$.

A higher minimum distance is desirable because it increases the capability to detect and correct errors. If the Hamming distance between codewords is very small, even minor errors can turn one valid codeword into another, making it impossible to detect an error. The goal of an encoder is to maximize the minimum distance for a given $n$ and $k$. In fact, given $n$ and $k$ the maximum achievable minimum distance is bounded by~\cite{mackay2003information}:

\begin{equation}
    d_{\min} \le n - k + 1
\end{equation}

Increasing $n$ generally helps achieve a larger minimum distance, which improves error detection and correction. However, a larger $n$ also means that each codeword occupies more bits, requiring increased number of channel use to send a message, and therefore slowing down the communication. 

We briefly introduce a simple and widely used class of error-correcting codes known as linear codes. In a linear code, the encoding function is a linear transformation represented by a matrix of the form:
\begin{equation}
    \E^T = 
\begin{bmatrix}
I_k\\
P
\end{bmatrix},
\end{equation}
where $I_k$ is the $k \times k$ identity matrix and $P$ is a $(n-k) \times k$ matrix known as the parity matrix. Given an input message $\mathbf{s} \in \{0,1\}^k$, represented as a column vector, the corresponding codeword $\mathbf{t} \in \{0,1\}^n$ is obtained by:
\begin{equation}\label{eq:validcodes2}
    \mathbf{t} = \E^T \mathbf{s}.
\end{equation}
The first $k$ bits of the codeword are a direct copy of the input, while the remaining $n-k$ bits—called parity bits—are linear combinations of the input bits as specified by $P$. These parity bits encode redundancy that enables error detection and correction. A commonly used decoding strategy for linear codes is the syndrome decoder.

Syndrome decoding relies on a key property of linear codes known as the syndrome. The syndrome is a binary vector associated with each codeword: it is zero if the word is a valid codeword, and non-zero if an error has occurred during transmission. Moreover, the value of the syndrome often contains information about the location of the error, making it useful for correction. For a linear code $\E = [I_{k\times k}, P^T]$, the corresponding \textit{parity check matrix} $\Hh$ is defined as

\begin{equation}
    \Hh = 
    \begin{bmatrix}
    P &I_{n-k}
    \end{bmatrix}.
\end{equation}
The syndrome associated with a received codeword $\mathbf{r} \in \{0,1\}^n$ is defined as $\mathbf{z} = \Hh \mathbf{r}$, where $\Hh $ is the parity-check matrix. A key property of linear codes is that all valid codewords lie in the kernel of $\Hh$, i.e.,

\begin{equation}\label{eq:validcodes}
    \Hh \mathbf{t} = \mathbf{0}, \quad \text{for all valid codewords } \mathbf{t} = \E^T \mathbf{s}.
\end{equation}
Thus, any received codeword with a non-zero syndrome indicates that an error has occurred during transmission. The syndrome not only detects the presence of an error but can also be used to locate it. Let $\mathbf{e}$ denote the binary error vector introduced by the channel. Then the received string $\mathbf{r}$ can be expressed as:
\begin{equation}
\mathbf{r} = \mathbf{t} \oplus \mathbf{e},
\end{equation}
where $\mathbf{t}$ is the transmitted codeword and $\oplus$ indicates a modulo 2 addition. Using \eqref{eq:validcodes}, the syndrome of the received string becomes:
\begin{equation}
\mathbf{z} = \Hh \mathbf{r} = \Hh(\mathbf{t} \oplus \mathbf{e}) = \Hh \mathbf{e}.
\end{equation}
This shows that the syndrome depends solely on the error vector \( \mathbf{e} \), and not on the original codeword \( \mathbf{t} \), and this can be used to detect and correct error. The third equality, $\mathbf{z} = \Hh\mathbf{e}$, establishes a relationship between the syndrome of a received codeword and the error vector $\mathbf{e}$ that might have occurred during transmission. With a precomputed lookup table that maps all possible error vectors $\mathbf{e}$ to their corresponding syndromes $\mathbf{z} = \Hh\mathbf{e}$, one can effectively decode the received message by (i) observing the syndrome $\mathbf{z}$ of the received vector $\mathbf{r}$, and then (ii) using the lookup table to identify the most probable error vector $\Tilde{\mathbf{e}}$ corresponding to the observed syndrome $\mathbf{z}$, and finally, (iii) predict the originally transmitted valid codeword by adding this error vector (modulo 2) to the received codeword $\mathbf{r}$.
\begin{equation}
    \hat{\mathbf{t}} = \mathbf{r}\oplus\Tilde{\mathbf{e}}
\end{equation}

It is important to emphasize that despite all the syndrome decoding, there's always a non-zero possibility of making an incorrect prediction of $\tilde{\mathbf{e}}$ 
since there could be more than one $\mathbf{e}$ associated with a single $\mathbf{z}$. In that case, the decoder has to make the most probable estimate of $\mathbf{e}$. An incorrect estimate $\tilde{\mathbf{e}}$ leads to a decoding error. The block error rate is the probability that an entire transmitted codeword (or block) is decoded incorrectly, i.e., $P(\hat{\mathbf{t}} \ne \mathbf{t})$. Unlike bit error rate, which measures errors per individual bit, block error rate captures whether the decoder successfully recovers the full original message block without any errors. Reducing the block error rate generally requires increasing the redundancy of the code by making the codewords longer, which also consequently leads to a decrease in the communication rate~\cite{mackay2003information}. 

\subsection{Relevant stochastic thermodynamics}\label{stoch_thermo}

Equilibrium statistical mechanics derives thermodynamics properties of a system from the underlying equilibrium statistics. Stochastic thermodynamics is an extension which defines thermodynamic quantities such as heat, work, and entropy to the level stochastic trajectories in processes evolving arbitrarily far from equilibrium. The key assumption in stochastic thermodynamics is that any degree of freedom not explicitly described by the system's dynamics—such the heat or particle reservoirs remains in equilibrium. This allows us to connect thermodynamic quantities and dynamics, through the principle of local detailed balance~\cite{seifert2012stochastic, seifert2018stochastic, esposito2010entropy, esposito2012stochastic}. 

One of the central quantities in stochastic thermodynamics is entropy production (EP), which quantifies the irreversible heat dissipation during a process. Consider a system with a discrete state space $\mathcal{X}$ that evolves over a time interval $[0, \tau]$, transforming an initial distribution $p_{X_0}$ into a final distribution $p_{X_\tau}$. Suppose the system is coupled to several heat reservoirs indexed by $i$, each characterized by an inverse temperature $\beta_i$. Let $\mathcal{Q}_i(x_0)$ denote the expected heat flow from the system into the $i^{\text{th}}$ reservoir, conditioned on the system starting in state $x_0$. The expected total heat flow into the $i^{\text{th}}$ reservoir over the process is given by:

\begin{equation}
    \langle \mathcal{Q}_i \rangle_{p_{X_0}} = \sum_{x \in \mathcal{X}} p_{X_0}(x) \mathcal{Q}_i(x).
\end{equation}
With this, the total entropy production during the process is expressed as:

\begin{equation}\label{EP_def}
    \C(p_{X_0}) = \sum_{i} \beta_i \langle \mathcal{Q}_i \rangle_{p_{X_0}} - \left[ S(p_{X_0}) - S(p_{X_\tau}) \right],
\end{equation}
where $S(p) = -\sum_x p(x) \log p(x)$ is the Shannon entropy. The first sum in the expression is also called the \textit{total entropy flow} and it corresponds to the change in the thermodynamic entropy of the environments during the process. Furthermore, the total entropy flow from the system to the reservoirs can be written compactly as:

\begin{equation}
    \sum_i \beta_i \langle \mathcal{Q}_i \rangle_{p_{X_0}} = \sum_{x \in \mathcal{X}} p_{X_0}(x) f(x) = \langle f \rangle_{p_{X_0}},
\end{equation}
where $f(x) := \sum_i \beta_i \mathcal{Q}_i(x)$ is the expected change in the thermodynamic entropy of all the reservoirs combined if the system starts in state $x$.

Thus, $\C(p_{X_0})$ is the part of the total entropy flow that is irreversibly lost into the environment, entropy production, whereas $\left[S(p_{X_0}) - S(p_{X_1})\right]$, which is called Landauer cost is the reversible part of the total heat flow. According to the second law of thermodynamics, entropy production is always non-negative:
\begin{equation}
    \C(p_{X_0}) \ge 0.
\end{equation}
The expected entropy flow $\langle f \rangle_{p_{X_0}}$, which captures the average heat dumped into the reservoirs, is therefore always lower bounded by the Landauer cost. This lower bound is saturated in the limit of quasi-static (infinitely slow) or near-equilibrium processes, where $\C(p_{X_0}) \to 0$. Moreover, the Landauer cost itself is upper bounded by $\ln |\mathcal{X}|$. In contrast, for processes that are far from equilibrium, the total entropy flow $\langle f \rangle_{p_{X_0}}$ can far exceed this bound, meaning that EP dominates the total thermodynamic cost.

\begin{figure*}
    \centering
    \includegraphics[trim = {0 8cm 0 0}, width=1\linewidth]{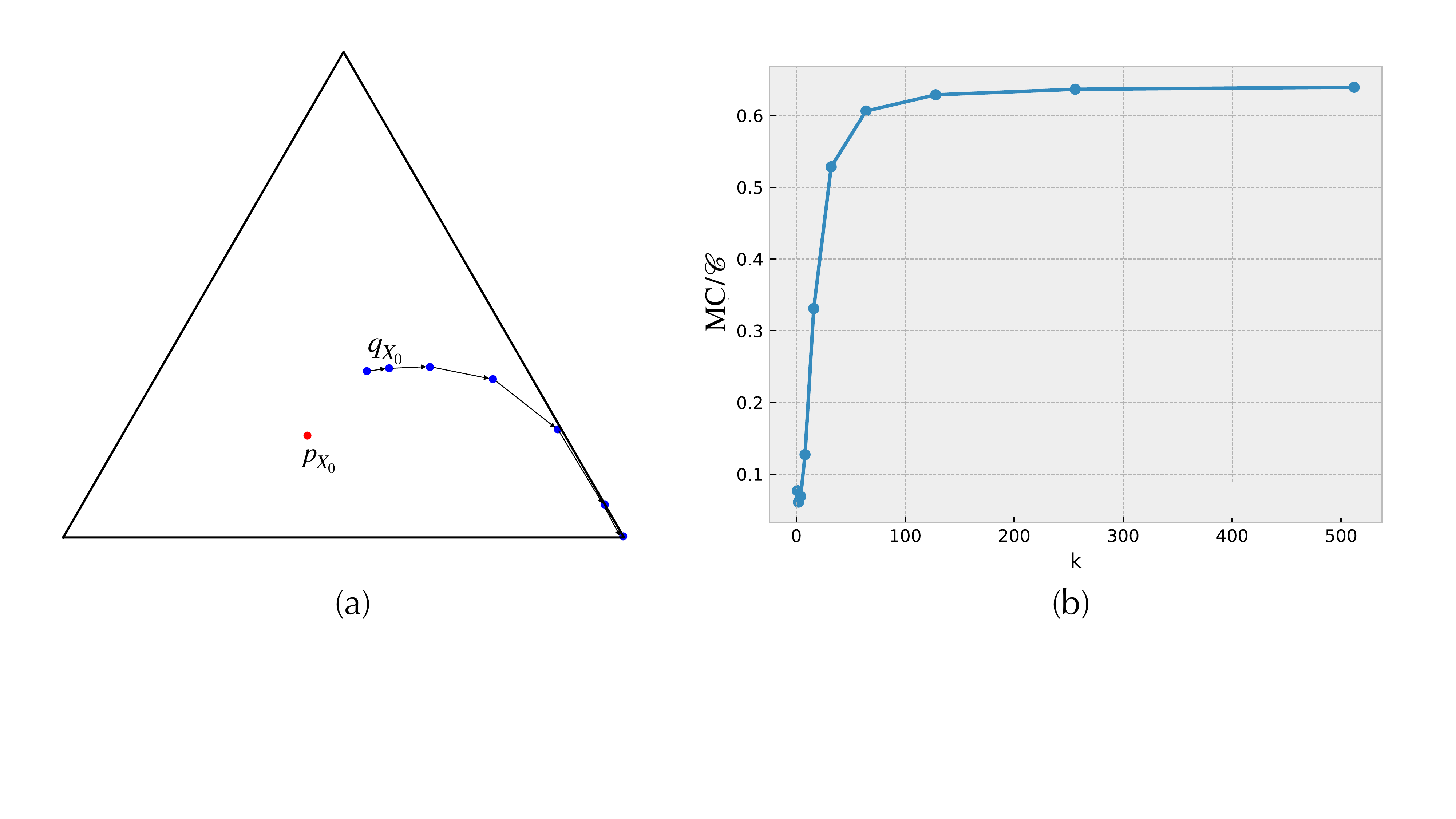}
    \caption{
    Mismatch cost increases significantly with entropy flow scaling $k$, demonstrating its dominant contribution in high-EP scenarios. Starting with arbitrary chosen values $f(1) = 0.5$, $ f(2) = 0.1 $, and $ f(3) = 0.2 $, we scale all $ f(x) $ by a factor $k$, resulting in new values  $k f(x)$.  
    (a) As $k$ increases, the associated prior distribution $q_{X_0} = \mathrm{arg}\min_{r_{X_0}} \C(r_{X_0})$ (as defined by Eq.~\ref{EPdef2}) shifts closer to the boundary of the 2-simplex.  
    (b) Consequently, the mismatch cost for a typical initial distribution (here, $ p_{X_0} = \{0.46, 0.33, 0.21\} $) becomes a larger fraction of the total EP, eventually exceeding $60\%$ of the total EP. The calculations are performed using the map $G$ defined in Eq.~\ref{eq:mapG}, with $\phi$ set to 0.1. The total entropy production is computed from $f(x)$ and $G$ using Eq.~\ref{EPdef2}, while the corresponding mismatch cost for each resulting prior is calculated using Eq.~\ref{eq:MMCdef}.}
    \label{fig:simplex_MMC}
\end{figure*}

Advances in stochastic thermodynamics have led to the discovery of various contributions to total EP based on key properties of a system's dynamics. Notably, thermodynamic uncertainty relations (TURs) establish a contribution to EP that arises from the precision of currents involved in the system's evolution~\cite{barato2015thermodynamic, gingrich2016dissipation, horowitz2020thermodynamic}. Another significant class of results, known as speed limit theorems (SLTs), provide a lower bound on EP in terms of the average number of state transitions during the system’s evolution~\cite{shiraishi2018speed, vo2020unified}. These results require some key details of systems dynamics to lower bound EP, for instance, SLTs require knowledge of the average number of transitions in the underlying dynamics, while TURs depend on the precision of currents involved in the process.

A distinct and complementary contribution to EP is known as the mismatch cost. Consider the initial distribution that minimizes EP in Eq.~\ref{EP_def}, denoted as $q_{X_0}$, which we refer to as the prior distribution of the process. When the process starts in a distribution different from $q_{X_0}$, the resulting excess EP is given by the mismatch cost. 
$\C(p_0)$, can be decomposed as~\cite{kolchinsky2017dependence, kolchinsky2021dependence},

\begin{equation}\label{eq:MMC1}
    \C(p_0) = [D(p_{X_0} \| q_{X_0}) - D( p_{X_\tau} \| q_{X_\tau})] + \C_{\text{res}},
\end{equation}
where $D(p \| q)$ is the Kullback-Leibler (KL) divergence between $p$ and $q$, and $\C_{\text{res}}$, known as the residual EP, represents the minimum EP of the process, defined as $\C_{\text{res}} := \C(q_0)$. The drop in KL-divergence $D(p_{X_0} \| q_{X_0}) - D( p_{X_\tau} \| q_{X_\tau})$ is called the mismatch cost (MMC). Moreover, the mapping of any initial distribution of system's state to final distribution under a dynamics can be compactly described by a stochastic map $G$ such that $p_{X_\tau}(x) = \sum_{x' \in \X} G(x|x')p_{X_0}(x')$, or $p_{X_{\tau}} = Gp_{X_0}$. Then, MMC can be written as
\begin{equation}\label{eq:MMCdef}
    \MMC(p_{X_0}) = D(p_{X_0} \| q_{X_0}) - D( Gp_{X_0} \| Gq_{X_0})
\end{equation}
Due to the data processing inequality for KL divergence, MMC is always non-negative~\cite{polyanskiy2014lecture}. By the second law, the residual EP satisfies $\C_{\text{res}} \ge 0$, and consequently MMC provides a non-negative lower bound on total cost. The detailed physical implementation of the process is encoded in the residual cost and the prior distribution. However, once the prior is known, the MMC gives a guaranteed lower bound on EP for any arbitrary initial distribution. Crucially, this bound applies broadly to a general Markovian or non-Markovian process.

Furthermore, in~\cite{yadav2024mismatch}, it has been shown that MMC can be a significant contribution to the total cost, especially when total heat generation during the process is significant. To sketch the intuition: recall that $f(x)$ represents the expected entropy flow to the reservoirs when the system starts in state $x$. One can re-write Eq.~\ref{EP_def} as

\begin{equation}\label{EPdef2}
    \C(p_{X_0}) = \sum_{x} f(x) p_{X_0}(x) - \left[ S(p_{X_0}) - S(Gp_{X_0})\right]
\end{equation}
Prior distribution is entirely determined by $f(x)$ and the map $G$~\cite{yadav2024mismatch}. In a scenario, where the total heat generation is large on a scale of $\ln|\X|$%, i.e., when $f(x) \gg \ln|\mathcal{X}|$ 
and $f(x)$ is not uniform across all states $x \in \X$, the resulting prior distribution  gets closer to the edges of the probability simplex. In other words, the prior becomes increasingly peaked. This peaking of the prior results in a high value of KL divergence between a typical actual distribution $p_{X_0}$ and the prior $q_{X_0}$, thereby increasing the MMC.  More precisely, if $\max_{x \in \mathcal{X}} f(x) - f(c) > \ln |\mathcal{X}|$, then mismatch cost for the worst-case input distribution, denoted as $\MMC^* = \max_{p_{X_0}} \MMC(p_{X_0}))$, is lower bounded by~\cite{yadav2024mismatch},
\begin{equation}
    \MMC^* \ge \max_{x} f(x) - \min_{x} f(x) - \ln |\X|.
\end{equation}

Figure~\ref{fig:simplex_MMC} illustrates this effect for a three-state system under a map $G$, 
\begin{equation}\label{eq:mapG}
    G = \begin{bmatrix}
        1-\phi & \phi & 0 \\
        0 & 1-\phi & \phi \\ 
        \phi& 0 & 1-\phi
        \end{bmatrix}.
\end{equation}
where $\phi \in [0, 1]$.
Starting from an arbitrary function $f(x)$, scaling its values by a factor of $k$ causes the prior distribution to shift toward the edges of the 2-simplex. As this scaling increases, the associated MMC becomes an increasingly significant contributor to the total EP for a typical initial distribution, as shown in Fig.~\ref{fig:simplex_MMC}.

In the context of communication channels, MMC is central to understanding how the rate of information transmission relates to its cost. As discussed earlier, the information transmitted from $X$ to $Y$ during each copy operation is quantified by the mutual information $I(X;Y)$, which depends on the input distribution $p_X$ for a fixed channel. From the decomposition of total thermodynamic cost in Eq.~\ref{eq:MMC1}, the MMC is the component of the cost that explicitly depends on $p_X$, while the residual term remains fixed across all input distributions. The input distribution to a communication channel determines the initial joint distribution over the channel's state. Since MMC captures the part of EP that depends on this initial distribution, it effectively isolates the component of EP that varies with the input. Consequently, when analyzing how changes in information transmission affect the total thermodynamic cost, MMC identifies precisely the portion that covaries with mutual information.

%\Rfff{The thermodynamic cost the authors explore throughout the manuscript, the mismatch cost, is only a component of the total thermodynamic cost for a process. While it is indeed a non-negative lower bound on the total EP, it need not be most, or even a significant portion, of the total thermodynamic cost. The authors note at the end of section II that the “mismatch cost can constitute a major portion of the total EP”, citing Ref 33, but make no argument for whether that should be the case in the processes they consider in this manuscript. The authors do not calculate the total EP of any of the processes considered in this manuscript. Given that the subject of the paper is the thermodynamic costs of communication, and that many of their main results involve the dependence of thermodynamic costs on various features of communication channels, this is a significant limitation. I would encourage the authors to compute the total EP in some of their example systems (where possible), with the goals of A) verifying that the mismatch cost is indeed a significant contributor to the total EP, and B) providing stronger evidence for their conclusions about the dependence of thermodynamic costs on features of communication processes.}

\begin{table*}
\begin{center}
\begin{tabular}{ |p{4cm}||p{13cm}|}
 \hline
 \multicolumn{2}{|c|}{Table of notation} \\
 \hline
 \centering Symbol & Definition \\
 \hline
  \centering $\pi_{Y|X}$& Conditional distribution representing the noise in the channel \\
   \hline
  \centering $p_X$& Input or source distribution. \\ 
   \hline
  \centering $p^a_{XY}$& Distribution over joint state of channel before copy operation.\\ 
   \hline
   \centering $p^b_{XY}$& Distribution over joint state of channel after copy operation.\\ 
   \hline
   \centering $q^a_{XY}$& Prior distribution of process of overwriting of input node.\\ 
   \hline
   \centering $q^b_{XY}$& Prior distribution of process of copying input state to output node.\\ 
   \hline
   \centering $\q^a_{XY}$& Distribution resulting from the prior $q^a_{XY}$ after overwriting of input node.\\ 
   \hline
   \centering $\q^b_{XY}$& Distribution resulting from the prior $q^b_{XY}$ after copy operation.\\ 
   \hline
   \centering $\MC^a$& Mismatch cost associated with the process of overwriting of input node. \\ 
   \hline  
   \centering $\MC^b$& Mismatch cost associated with the process of copying of input state to output node. \\ 
   \hline  
\end{tabular}
\end{center}
\caption{Table of notation for the main symbols used in the paper.} \label{TABLE}
\end{table*}

\section{Minimal thermodynamic cost in communication channels}
\label{thermo_cost_channels}

The model of communication channel that we are going to consider in this section is inspired by many real-world channels in which the source end is continuously and repeatedly updated with new values, and each of which is copied to the other end of the channel, the receiver end. Crucially, it is not a one-time process but it is continuously repeated---source values are repeatedly updated and copied to the receiver in an ongoing cycle. Such dynamics arise in a variety of communication channels---for example, in cellular signaling pathways, where ligand concentrations at the membrane are continuously sensed and propagated downstream~\cite{ouldridge2017thermodynamics, ten2016fundamental} and in digital devices, where sensors or processors continuously write new values to shared memory or communication buses for retrieval by downstream components.

\begin{figure}
    \centering
    \includegraphics[trim = {0 9cm 0 0}, width=\linewidth]{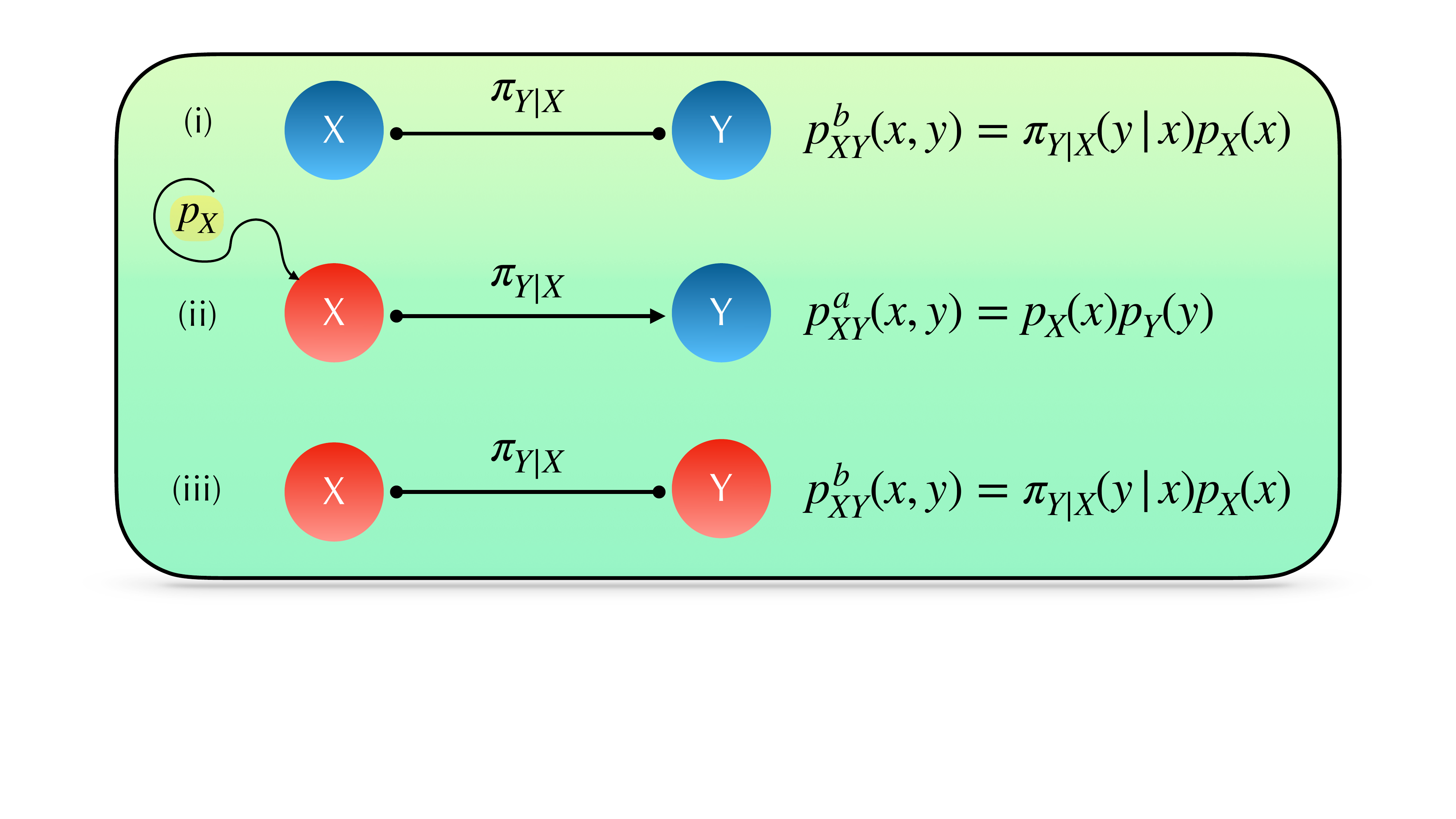}
    \caption{We model the communication channel as consisting of an input node $X$ and an output node $Y$. At each iteration, a value is sampled from the input distribution $p_X$ and copied to the output according to the conditional distribution $\pi_{Y|X}$. Following this copy operation, the input is updated with a new value, independently drawn from $p_X$, thereby resetting the joint state of $(X, Y)$ from correlated to uncorrelated. }
    \label{fig:model_and_examples}
\end{figure}

We represent a noisy communication channel as a two node graph with an input node $X$ and an output node $Y$, and a conditional distribution $\pi_{Y|X}$ that specifies the probability of output given input (see Fig.~\ref{fig:model_and_examples}). The joint distribution $p_{XY}$ defines the state of the channel. Communication involves two key steps: 
\begin{enumerate}
    \item[(a)] Overwriting of input: A value is assigned to the input node $X$, drawn randomly from a distribution $p_X$.
    \item[(b)] Copying input value to output: The input value is then copied to the output node $Y$, subject to noise governed by $\pi_{Y|X}$.
\end{enumerate} 
\noindent
Importantly, as in many practical scenarios where channels are repeatedly reused, in this model, steps (a) and (b) are applied in succession over and over again. More specifically, after the end of step (b) in each iteration of the channel, the output and input node are correlated. Let $p^b_{XY}$ denote the joint state of the channel at the end of step (b) from a previous cycle. It can be written as,
\begin{equation}\label{eq:pb}
    p^b_{XY}(x, y) = \pi_{Y|X}(y|x) \, p_X(x).
\end{equation}

For subsequent use of the channel, input node is overwritten with a new value that is randomly sampled from the distribution $p_X$, Let $p^a_{XY}$ denote the joint state of the channel after overwriting process (a). Since the new input value is sampled randomly, $p^a_{XY}$ is given by,
\begin{equation}\label{eq:pa}
      p^a_{XY}(x, y) = p_X(x) \, p_Y(y)
\end{equation}
where $p_Y(y) = \sum_x p^b_{XY}(x, y)$ is the marginal distribution over $Y$. The next step, (b), involves copying the new input value of $X$ to the output node $Y$ subject to the noise $\pi_{Y|X}$. After this copy process, the joint distribution returns to $p^b_{XY}(x, y)$.

In this way, each iteration of sending a bit through the channel causes the system to cycle from the state $p^b_{XY}$ to $p^a_{XY}$, and then back to its initial form $p^b_{XY}$. 

\subsection{Mismatch cost in communication channel}

The joint state of the channel consists of the input node $X$ and the output node $Y$. The process of input overwriting can be viewed as the evolution of this joint system, where the state of $X$ changes independently of $Y$, while the state of $Y$ remains unchanged. This type of evolution is known as a subsystem process (see App~\ref{App1}). In such processes, evolution of a joint system can be broken down into independent evolution of multiple subsystems~\cite{wolpert2019stochastic, wolpert2020uncertainty}. Importantly, in subsystem processes the prior distribution over the joint system factorizes into the product of the marginals of the two subsystems. 
Specifically, let $q^a_{XY}$ denote the prior for the input-overwriting process. Then it can be expressed as a product distribution:
\begin{equation}\label{eq:prior_ow}
q^a_{XY}(x, y) = q^a_X(x) q^a_Y(y).
\end{equation}
The new value of $X$ is independently sampled from $p_X$. If the channel starts in $q^a_{XY}$ before the input node is overwritten, the after the overwriting, the distribution of $X$ is $p_X$ and the resulting joint distribution is
\begin{equation}
    \q^a_{XY}(x, y) = p_X(x) q^a_Y(y).
\end{equation}

\begin{figure}
    \centering
    \includegraphics[width=0.6\linewidth]{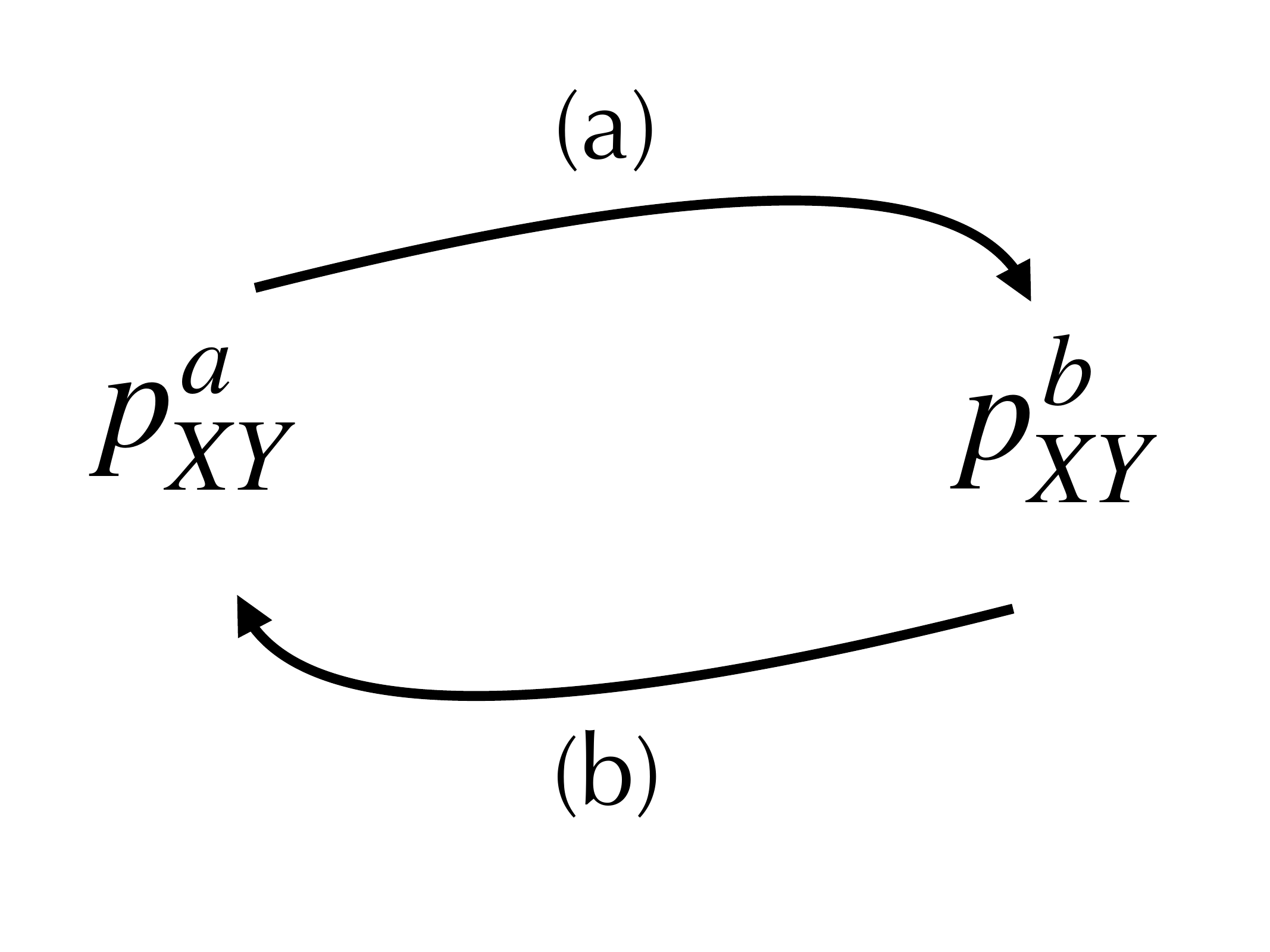}
    \caption{Diagram illustrating how the joint distribution of the channel evolves during (a) the overwriting of the input and (b) the noisy copying of the input state to the output. Here, $p^a_{XY}(x, y) = p_X(x)p_Y(y)$ and $p^b_{XY}(x, y) = \pi_{Y|X}(y|x)p_X(x)$. With each use of the channel, the joint distribution cycles from $p^b_{XY}$ to $p^a_{XY}$ and back again.}
    \label{fig2}
\end{figure}

\begin{figure}
    \centering
    \includegraphics[width=0.6\linewidth]{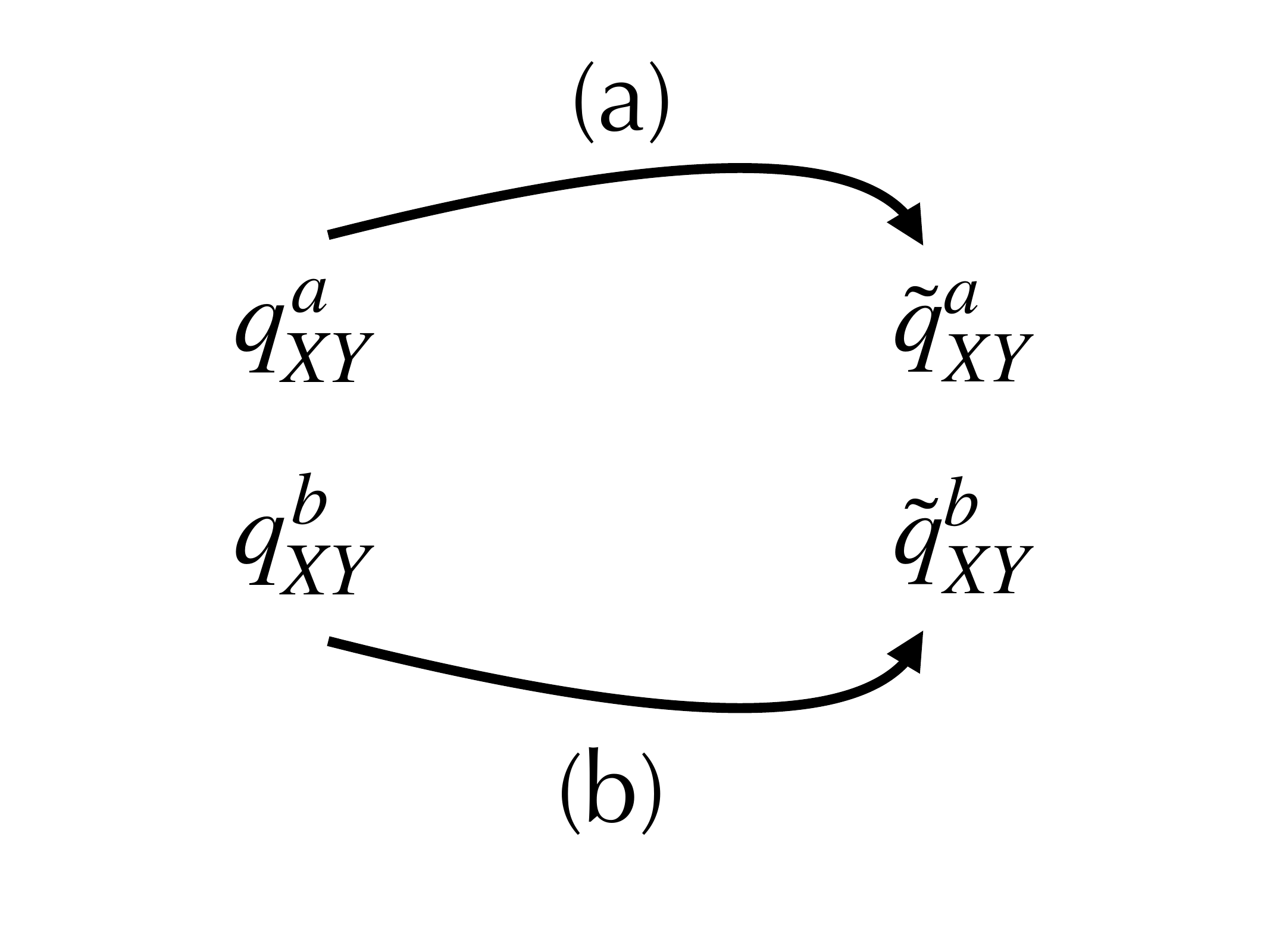}
    \caption{Diagram illustrating how the prior distributions associated with (a) the overwriting process and (b) the noisy copying of the input state to the output evolve through their respective transformations. In (a), the prior changes from $q^a_{XY}(x, y) = q^a_X(x) q^a_Y(y)$ to $\q^a_{XY}(x, y) = p_X(x) q^a_Y(y)$. In (b), the prior changes from $q^{b}_{XY}(x, y)$ to $\q^b_{XY}(x, y) = \pi_{Y|X}(y|x)q^b_X(x)$.}
    \label{fig3}
\end{figure}
The actual distribution over the states of the channel before and after the overwriting process are $p^b_{XY}$ (Eq.~\ref{eq:pb}) and $p^a_{XY}$ (Eq.~\ref{eq:pa}), respectively, while the prior distribution changes from $q^a_{XY}$ to $\q^a_{XY}$ as depicted in Fig.~\ref{fig3}. Therefore, the mismatch cost associated with this overwriting process is given by:
\begin{equation}
\MC^{a}(p_X) = D\left(p^b_{XY}\| q^a_{XY}\right)- D\left(p^a_{XY}\| \q^a_{XY}\right)
\end{equation}
which simplifies to,
\begin{equation}\label{eq:owMC}
\MC^{a}(p_X) = \I(X; Y) + D(p_X\| q^a_X).
\end{equation}
where $\I(X;Y)$ is the mutual information between $X$ and $Y$ before overwriting.

\begin{equation}
\I(X; Y) = \sum_{x, y \in X} p^b_{XY}(x, y) \log \frac{ p^b_{XY}(x, y)}{p_X(x) p_Y(y)}.
\end{equation}
The derivation is provided in App.~\ref{App2}.

Now, let us consider the mismatch cost associated with copying process. 
The value of the output depends on the value of the input $\pi_{Y|X}$. Therefore, although $Y$ evolves while the state of $X$ does not change, by definition it is not a subsystem process and the prior distribution cannot be expressed as a product distribution. Let $q_{XY}^b$ denote some arbitrary joint prior distribution for the copying process. After the copy operation, $q_{XY}^b$ evolves to 
\begin{equation}
    \q^b_{XY}(x, y) = \pi_{Y|X}(y|x)q^b_X(x),
\end{equation}
where $q^b_X(x) = \sum_y q_{XY}^b(x, y)$ is the marginal prior distribution of $X$. 
The actual distribution over the states of the channel before and after the copying process is $p^a_{XY}$ (Eq.~\ref{eq:pa}) and $p^b_{XY}$ (Eq.~\ref{eq:pb}) respectively.  Therefore, the mismatch cost associated with copying operation is

\begin{equation}
    \MC^b = D(p^a_{XY}\|q^b_{XY}) - D( p^b_{XY}\|\q^b_{XY}).
\end{equation}
which further simplifies to 
\begin{equation}\label{eq:52}
    \MC^b = D(p_Xp_Y\|q_{XY}^b) - D(p_X\|q_X^b).
\end{equation}
Combining (\ref{eq:owMC}) and (\ref{eq:52}), the total mismatch cost is,

\begin{align}
    \MC(p_X) %&= \MC_A+\MC_B \\
    &= \I(X;Y) + D(p_X\|q_X^a) \nonumber \\ 
    &\qquad + D(p_Xp_Y\|q_{XY}^b) - D(p_X\|q_X^b) \label{eq:MCvsMI}
\end{align}
\noindent
While $\I(X;Y)$ quantifies the rate of information transmitted per channel use, $\MC(p_X)$ in Eq.~\ref{eq:MCvsMI} expresses the MMC per use of the communication channel, representing the unavoidable EP incurred with each symbol transmission.

The derivation of Eq.~\ref{eq:MCvsMI} does not rely on any specific assumptions about the physical details of the communication channel’s implementation. In particular, it holds regardless of whether the underlying dynamics of steps (a) or (b) are Markovian or non-Markovian, discrete-time or continuous-time. The result follows directly from the general structure of the communication model, which consists of two fundamental operations: (a) overwriting the input (a subsystem process) and (b) performing a noisy copy to the output.

The only model-dependent quantities in Eq.~\ref{eq:MCvsMI} are the prior distributions, $q^a_X$ and $q^b_{XY}$, which encapsulate the physical characteristics of the process. As described earlier in Eq.~\ref{EPdef2}, these priors are determined by the entropy flow functions $f(x)$ and the computational map $G$. Once $G$ is specified, assumptions about the entropy flow $f(x)$ allow one to determine the corresponding priors. In the next section, we illustrate this by deriving priors from given entropy flow profiles.

Furthermore, since KL-divergence is always non-negative, $D(p_X \| q^a_X) \ge 0$, and by the data processing inequality, $D(p_X p_Y \| q^b_{XY}) - D(p_X \| q^b_X) \ge 0$ for all $p_X$, $q^a_X$, and $q^b_{XY}$. Therefore, 
\begin{equation}\label{MMC_MI_lower}
\MC(p_X) \ge \I(X; Y).
\end{equation}
This inequality holds irrespective of the prior of the underlying process implementing the communication channel. The total cost in the communication channel is lower bounded by the mutual information, which quantifies the rate of information transmission. 

\subsection{Implications for work extraction via measurement and feedback}

We briefly talk about the role of the minimal cost of communication in Eq.~\ref{MMC_MI_lower} to the work extraction under measurement and feedback control. 

A generalized maximum work extraction formulation of the second law states that in an isothermal process the maximum extractable work between two distributions $p_{X_0}$ and $p_{X_1}$ is given by the change in non-equilibrium free energy~\cite{ hasegawa2010generalization},
\begin{equation}\label{eq_feedback_control1}
    \avg{W} \le \F(p_{X_0}) -  \F(p_{X_1}),
\end{equation}
where $\F(p_{X_0}) = \avg{E}_{p_{X_0}} - k_B T S(p_{X_0})$. However, in the presence of measurement and feedback protocols, i.e., when there is a measurement of system's state is involved before applying the protocol and the protocol depends on the outcome of the measurement, the maximum extractable work is the change in free energy plus the mutual information between system's state and the memory~\cite{sagawa2012nonequilibrium, sagawa2010generalized}. 

\begin{equation}\label{eq_feedback_control2}
    \avg{W} \le \F_{E_0}(p_{X_0}) - \F_{E_1}(p_{X_1}) + \I (X; M)
\end{equation}
where $\I(X;M)$ is the mutual information between system's state and the memory of the measuring device. If the accuracy of a measurement is described by the conditional distribution $\pi_{M|X}$, which gives the probability of observing outcome $M$ given the system's state $X$, then the measurement process can be viewed as a communication (or observational) channel $\pi_{M|X}$. Equation~\ref{eq_feedback_control2} implies that it is possible to extract work exceeding the bound imposed by the Second Law (as given in Eq.~\ref{eq_feedback_control1}) by performing a measurement on the system and applying a feedback protocol that depends on the measurement outcome.

A canonical example of this phenomenon is Szilard’s engine, which consists of a single-particle gas in a box. In this setup, a demon measures whether the particle is on the left or right side of a partition. Based on this information, the demon allows the piston to expand, extracting an amount of work equal to $k_B T \log 2$~\cite{szilard1964decrease, sagawa2012thermodynamics}.

By repeating this cycle, the demon can extract $k_B T \log 2$ of work per iteration—or more generally, $k_B T \I(X; M)$, where $\I(X; M)$ is the mutual information between the system and measurement outcomes. This seems to suggest the possibility of a perpetual motion machine of the second kind, apparently in violation of the Second Law.

Earlier efforts to reconcile Szilard's engine with the Second Law emphasized that, in a repeated setting, the step of resetting the demon’s memory to a blank state is logically irreversible and incurs an associated Landauer’s cost~\cite{bennett1987demons}. This cost precisely offsets the additional work extracted, thereby restoring consistency with the Second Law.  
More recent work has further deepened our understanding by showing that a loss of correlation between components of a system during a process contributes to the total EP. Specifically, the decrease in mutual information between subsystems during a process corresponds exactly to a positive contribution to the entropy production~\cite{wolpert2020uncertainty, ito2013information}. 

Consistent with prior work, Eq.~\ref{MMC_MI_lower} shows that there is a minimal thermodynamic cost associated with measurement in each cycle, and this cost is precisely given by the mutual information between the system’s state $X$ and the measurement device’s memory $M$. Furthermore, Eq.~\ref{eq:MCvsMI} provides an even tighter lower bound on the thermodynamic cost---one that can exceed the mutual information. Together, these results demonstrate that a measurement-and-feedback control process necessarily dissipates at least as much, and potentially more, work than it extracts. This reinforces the thermodynamic consistency of information-driven processes and precludes the possibility of a perpetual motion machine of the second kind.

In the next section, we examine a binary communication channel and begin by making assumptions about the associated priors. These assumptions allow us to analyze the channel's thermodynamic cost without needing to explicitly define the underlying physical process. We will demonstrate that, for certain kinds of priors, the observed relationship aligns with the concave relationship between EP and communication rate that's been found in concrete physical models~\cite{tasnim2024entropy, yan2024entropy}. However, we also show that this concavity is not universal, as other relationships can arise depending on the prior.

\subsection{Example of a binary channel}\label{example_binary_channel}

\begin{figure*}
  \includegraphics[trim = {0 11cm 0 0}, width=0.9\textwidth]{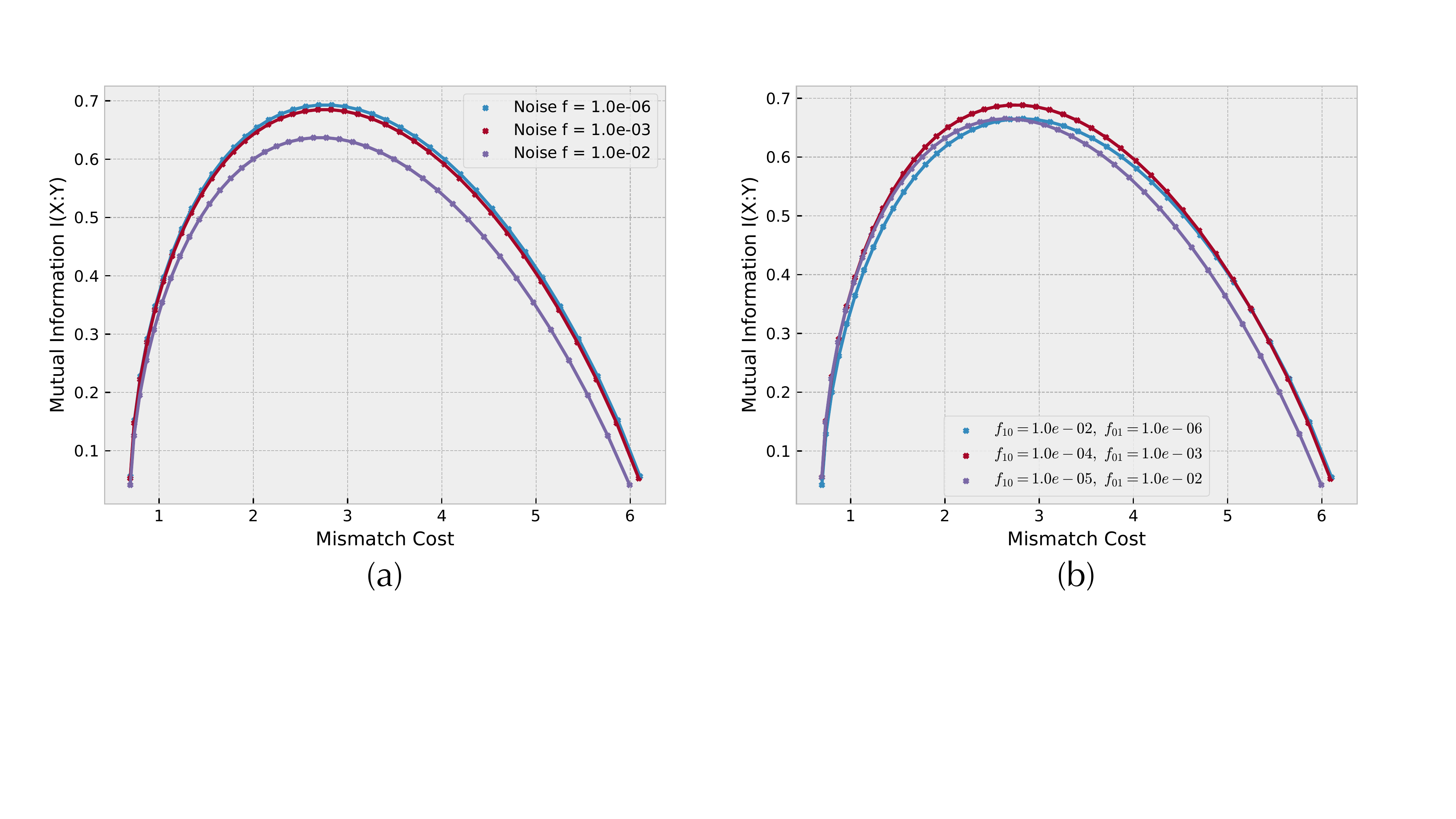}
    \caption{Relationship between mismatch cost and transmitted information in a binary channel. The graphs are obtained by using Eq.~\ref{eq:MCvsMI}. For the binary channel, the input distribution is parameterized as $p_X(x=0) = r$ and $p_X(x=1) = 1-r$. Each curve is generated by varying $r$ from $0$ to $1$, with the resulting mutual information between $X$ and $Y$ plotted on the y-axis and the corresponding total mismatch cost from Eq.~\ref{eq:MCvsMI} on the x-axis. The overwriting prior $q^a_X$ is taken to be uniform, while the prior for the copying process is set as $q^b_{XY}(0,0) = q^b_{XY}(0,1) = 0.002$ and $q^b_{XY}(1,0) = q^b_{XY}(1,1) = 0.498$. Distinct curves correspond to different noise levels in (a) symmetric channels and (b) asymmetric channels, parameterized by the conditional probabilities $\pi_{Y|X}(0|1) = f_{01}$ and $\pi_{Y|X}(1|0) = f_{10}$. In both cases, mutual information initially increases with mismatch cost, peaks at channel capacity, and then decreases. This concave-down relationship reflects diminishing returns: higher thermodynamic cost yields progressively smaller gains in information transmission.}
\label{fig:2}
\end{figure*}

We consider a binary communication channel with input and output state spaces $\X = \Y = \{0, 1\}$. For simplicity, we first assume that the prior distribution associated with the overwriting process, $q_{X}^a$ in Eq.~(\ref{eq:MCvsMI}), is uniform. We begin by assuming that the prior associated with the copying operation satisfies $q^b_{XY}(0, 0) = q^b_{XY}(0, 1) \ll q^b_{XY}(1, 0) = q^b_{XY}(1, 1)$. As shown in Fig.~\ref{fig:simplex_MMC}, such an asymmetric prior-positioned near the edge of the simplex—reflects large values of the entropy flow terms. A sufficient condition for obtaining this type of prior is a communication channel where $f(x, y)$ values are significantly large, with $f(0, 0) = f(0, 1) \ge f(1, 0) = f(1, 1)$. The source distribution $p_X(x)$ is parameterized by a single variable $r$, where $p_X(0) = r$ and $p_X(1) = 1 - r$. By varying $r$ over the interval $[0,1]$, we explore different source distributions. For each $p_X(x)$, we compute the mutual information between $X$ and $Y$ after the copy operation, and the mismatch cost using Eq.~(\ref{eq:MCvsMI}). 

Figure~\ref{fig:2} illustrates the relationship between mutual information and mismatch cost for both symmetric and asymmetric binary channels. 
Mutual information initially increases with mismatch cost, peaks at channel capacity, and then declines. 
This behavior arises because of how mutual information and mismatch cost vary with the input distribution parameter $ p_X(x=1) = r $. As $r$ varies from $0$ to $1$, the mutual information begins at zero, increases to a maximum at $r = r^*$ (with $r^* = 0.5$ for a binary symmetric channel), and then decreases again, reaching zero at $r = 1$. In contrast, the mismatch cost is minimized at an intermediate value of $r$ depending on the prior distribution, but maximized at the boundaries (either at $r = 0$ or $r = 1$). Thus, when $r$ approaches either boundary, the mutual information goes to zero while the mismatch cost is maximized. This behavior results in concave-down curves that reflect the law of diminishing returns: the increase in mutual information with increasing  thermodynamic cost gets progressively smaller.

It suggests that operating at channel capacity is thermodynamically inefficient. This principle---known as reverse multiplexing—shows that distributing cost across multiple channels can enhance overall information transfer efficiency~\cite{balasubramanian2015heterogeneity, balasubramanian2021brain}. 

\begin{figure}[h]
    \centering    \includegraphics[trim = {0 0 0 0}, width=1\linewidth]{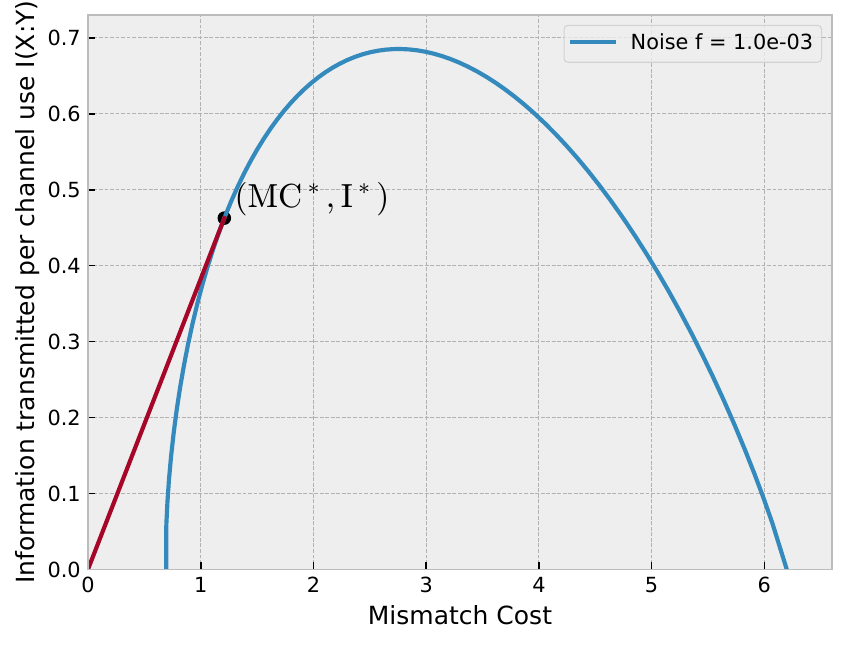}
    \caption{Relationship between information transmitted and mismatch cost per channel use is concave downwards, illustrating the law of diminishing returns. The point $(\MMC^*, \I^*)$ on the curve, determined by the tangent from origin, is where the transmission rate to cost ratio is maximum. }
    \label{fig:single_MCvsMI_symmetric}
    
\end{figure}

For example, Fig.~\ref{fig:single_MCvsMI_symmetric} shows an operating point $(\MC^*, \I^*)$ where the ratio of information transmitted to mismatch cost is maximized. Beyond this point, the curve exhibits diminishing returns—any further increase in information transmission rate $\Ii > \I^*$ requires disproportionately higher mismatch cost per additional bit of information. In such cases, it is more thermodynamically efficient to split the communication across two identical channels, $(X_1, Y_1)$ and $(X_2, Y_2)$, rather than using a single channel at a higher rate. If the values of inputs $X_1$ and $X_2$ are sampled independently from $p_{X_1}$ and $p_{X_2}$ respectively, the combined information transmitted across the two channels is simply the sum of the information transmitted through each channel individually:

\begin{equation}
\I(X_1, X_2; Y_1, Y_2) = \I(X_1; Y_1) + \I(X_2; Y_2).
\end{equation}
In Appendix~\ref{App3}, we provide a formal proof that the MMC associated with communication over a joint channel—comprising two input nodes and two output nodes—is equal to the sum of the MMC incurred when each of the two input-output channels operate independently. That is, if the values of input nodes are sampled independently from distributions $p_{X_1}$ and $p_{X_2}$, then the MMC incurred during communication over joint channel, $\MMC_{12}(p_{X_1}p_{X_2}) = \MMC_1(p_{X_1}) + \MMC_2(p_{X_2})$. This result demonstrates the additivity of MMC under parallel composition of communication channels.

Since each channel operates at or below $\I^*$, where the information-to-cost ratio is higher, the combined mismatch cost $\MMC_1 + \MMC_2$ will be lower than the mismatch cost required to transmit the same total information through a single high-rate channel. This reflects the principle of reverse multiplexing, where distributing the load across multiple lower-rate channels reduces the overall thermodynamic cost.

\begin{figure*}
    \includegraphics[trim = {0 10cm 0 0}, width=1\linewidth]{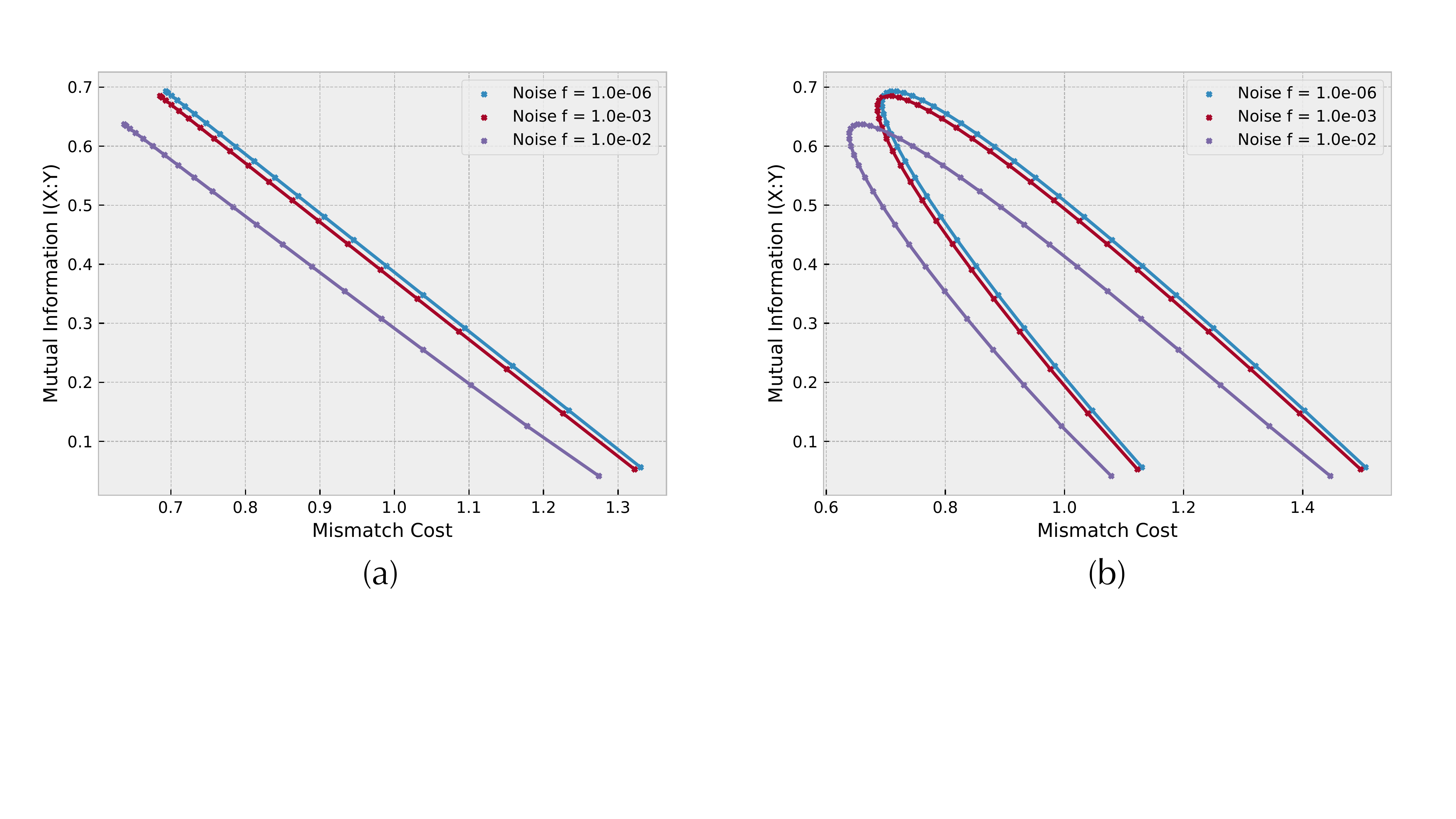}
    \caption{Relationship between mismatch cost and mutual information in a binary channel for different prior distributions.  
    (a) Both $q^a_X$ and $q^b_{XY}$ are uniform, leading to an inverse relationship: mismatch cost is minimized when mutual information is maximized (at channel capacity).  
    (b) $q^a_X$ remains uniform, but $q^b_{XY}$ is chosen asymmetrically with $q^b_{XY}(0, 0) = 0.35$, $q^b_{XY}(0, 1) = 0.22$, $q^b_{XY}(1, 0) = 0.25$, and $q^b_{XY}(1, 1) = 0.18$, resulting in a qualitatively different relationship. Once again, each curve is generated by varying input distribution parameter $p_X(x = 1) = r$ from $0$ to $1$, with the resulting mutual information between $X$ and $Y$ plotted on the y-axis and the corresponding total mismatch cost from Eq.~\ref{eq:MCvsMI} on the x-axis. 
    }
    \label{channel1}
\end{figure*}

However, this concave downward relationship does not hold for all choices of prior $q_{XY}^b$. For example, when both $q_{XY}^b$ and $q_X^a$ are uniform, the relationship between mutual information and mismatch cost is qualitatively different. As shown in Fig.~\ref{channel1}(a), mismatch cost and rate of information transmission becomes inversely related: MMC is minimized when the rate of information transmission reaches its maximum (i.e., channel capacity), indicating that operating at capacity is thermodynamically efficient in this case. Conversely, mismatch cost is highest when information transmitted is minimal. The reason behind this is that in a symmetric noisy channel, information transmission is maximized when the source distribution is uniform. On the other hand, when the priors $q_X^a$ and $q_{XY}^b$ are uniform, mismatch cost is minimized, since $D(p_X\|q_X^a) = 0$ and $D(p_{XY}\|q_{XY}^b)$ is minimum. As the source distribution deviates from uniformity, mutual information decreases while mismatch cost increases, yielding the inverse relationship shown in Fig.~\ref{channel1}(a).

Interestingly, the relationship can behave quite differently under other priors. Fig.~\ref{channel1}(b), generated with an arbitrary choice of $q_{XY}^b$ (while keeping $q_X^a$ uniform), shows a different relationship between mismatch cost and mutual information.

We now turn to the mismatch cost associated with encoding and decoding protocols in communication processes. As discussed earlier, encoding and decoding form the backbone of modern communication theory and practice. Thus, a comprehensive understanding of the thermodynamic cost of communication must also account for the energetic costs of encoding and decoding.

Unlike the simple yet broadly applicable model of a noisy communication channel considered earlier, encoding and decoding are inherently complex computational tasks. There exist multiple ways to implement them---for instance, linear encoders can be realized through Boolean circuits, while even basic implementations of syndrome decoders involve nontrivial circuit architectures. This complexity makes it challenging to account for the thermodynamic cost incurred at every computational step.

To circumvent these difficulties, we adopt the framework of periodic machines, as introduced in~\cite{yadav2024mismatch}. Periodic machines provide a way to implement algorithms at either a high-level or low-level abstraction, designed to reflect the periodic, clock-driven nature of modern computing architectures (e.g., RASP-based models). By design, periodic machines repeat a fixed physical process—analogous to the fetch-decode-execute cycles in stored-program computers. Importantly, these machines are tailored to the specific algorithm they implement. In the next section, we briefly review the properties of periodic mismatch cost and employ this framework to derive lower bounds on EP associated with encoding and decoding algorithms implemented on a periodic machine.

\section{Minimal thermodynamic cost of linear error correcting codes}\label{thermo_cost_codes}
\subsection{Mismatch Cost of Algorithmic Computation}\label{MC_of_Algorithms}

In this section, we formally define the concept of a periodic machine~\cite{yadav2024mismatch}, which iteratively executes the steps of a program. A program or algorithm is a sequence of instructions that systematically operate on variables, updating their values step by step as the computation progresses. All variables of an algorithm include input variables, internal variables such as flags, loop counters, and other control variables. A central component is the program counter, which tracks the current instruction being executed at any given moment. The state of the algorithm at any point during its execution is defined by the values of all its variables, including the current value of the program counter. Let $X_i$ denote the random variable representing the state of the algorithm immediately after the $i^\text{th}$ algorithmic step, with possible joint values $\x_i \in \X_\A$. Here, $\X_\A$ is the set of all allowed joint states of the variables in the algorithm $\A$. Additionally, $X_0$ represents the random variable corresponding to the initial state $\x_0$ of the algorithm before execution begins.

A jump from any state $\x_i$ to a next $\x_{i+1}$ is fully  specified by the algorithm. The transition is many-to-one or one-to-one in case of a deterministic algorithm, whereas in a probabilistic algorithm, this transition is stochastic, and the next state is drawn from a probability distribution specified by the algorithm. In both cases, the progression from $\x_i$ to $\x_{i+1}$ can be captured by a map $G_{\A} : \X_\A \to \X_\A $, corresponding to algorithm which, given the current state $\x_i$, returns the next state $\x_{i+1}$:
\begin{equation}
    \x_{i+1} = G_\A(\x_i)
\end{equation}
When the algorithm halts, $G_\A$ reaches a fixed point. The dynamics of the algorithm induce a corresponding dynamics on the probability distribution over the state space $\A$. To describe this formally, we index each state $\x_i \in \X_\A$ and define a transition matrix $G$, where each element $G_{nm}$ specifies the probability of transitioning from the state with index $m$ to the state with index $n$. If state $m$ does not transition to state $n$, set $G_{nm} = 0$. For deterministic algorithms, $G_{nm}$ will be either 0 or 1, reflecting the deterministic nature of the state transitions. The discrete-time evolution of the probability distribution over the state space is governed by the stochastic map $G$:

\begin{equation}\label{eq:DTMC}
    p_{X_{i+1}} = Gp_{X_{i}},
\end{equation}
where $p_{X_i}$ represents the probability distribution of $X_i$. Eq.~\ref{eq:DTMC} defines a discrete-time evolution of the probability distribution, where the same stochastic map $G$ is applied at each iteration. Therefore, one can write,
\begin{equation}\label{eq:36}
    p_{X_{i}} = G^i p_{X_{0}}.
\end{equation}
As the algorithm progresses and eventually halts, the distribution converges to a steady-state distribution. 

Assuming that the underlying physical process implementing $G$ remains unchanged across iterations, this corresponds to repeatedly applying the same physical process, driving the system through a sequence of probability distributions. Consequently, the prior distribution associated with the underlying process implementing the stochastic map $G$ also remains same across iterations. Let $q_X$ denote the prior. After each iteration of the process, the ending distribution associated with the prior is $\q_{X} = Gq_{X}$, whereas the actual distribution after $i^{th}$ iteration is given by Eq.~\ref{eq:36}. Therefore, the MMC in $i^{th}$ iteration is given by,
\begin{equation}\label{eq:PMC}
    \MC^i(p_{X_0}) = D(G^{i-1}p_{X_0} \| q_X) - D(G^i p_{X_0} \| Gq_X)
\end{equation}
for any $i\in\{0, 1, ...\}$. Note that the prior distribution $q_0$ remains same across iterations. However, the actual state distribution evolves with each iteration. Therefore, if the algorithm takes $r$ iteration to halt for all it's inputs, the total mismatch cost associated with an initial distribution $p_{X_0}$ is given by:
\begin{align}\label{eq_sum_MMC_periodic}
    \MC(p_{X_0}) &= \sum_{i = 1}^{r}\MC^i(p_{X_0}) \nonumber \\
    &= \sum_{i = 1}^{r} \left[D(G^{i-1}p_{X_0} \| q_X) - D(G^i p_{X_0} \| Gq_X)\right]
\end{align}

Even if the process starts at the prior $q_0$, after the first iteration, the distribution becomes $p_1 = G q_0 $, which differs from $q_0$. This deviation from the prior distribution results in a strictly positive MMC in the next iteration, and the same holds for subsequent iterations~\cite{wolpert2024thermodynamics}.

\subsubsection*{Initial state of the algorithm and resetting cost}\label{sec:resetting_cost}

Before going further, let us briefly discuss how the initial distribution for an algorithm is properly defined. When assigning an initial distribution $p_{X_0}$, it is crucial to understand the nature of the variables involved. An algorithm starts with its program counter set to 0 and includes input and non-input variables. Non-input variables may consist of loop counters, flags, and other special variables that are initialized to specific starting values. Let $X_{in}$, $X_{nin}$, and $X_{sp}$ denote the random variables representing the states of input variables, non-input variables, and special variables, with values $x_{in}$, $x_{nin}$, and $x_{sp}$, respectively. We assume that the input variables are freshly initialized before each run by sampling from a distribution $p_{X_{in}}$. The special variables—such as the program counter and loop counters—are always reset to their initialized states at the start of each run. This is represented by a deterministic distribution over their initial values:
\begin{equation}
    p_{X_{sp}}(x_{sp}) = \delta_{X_{sp}}(x_{sp}),
\end{equation}
where $\delta_{X_{sp}}(x_{sp})$ is the Kronecker delta, ensuring that the special variables are set to their predefined initial states with probability 1. In contrast, the non-input variables are assumed to retain their values from the end of the previous execution of the algorithm. That is,
\begin{equation}
p_{X_{nin}}(x_{nin}) = \sum_{x_{in}, x_{sp}} p_{X_r}(\x),
\end{equation}
where $p_{X_r}(\x)$ is the joint distribution over all variables at the end of the previous run. Thus, after reinitialization, the joint distribution over the algorithm's state is given by:
\begin{equation}
    p_{X_0}(\x) = p_{X_{in}}(x_{in}) \, \delta_{X_{sp}}(x_{sp}) \, p_{X_{nin}}(x_{nin}),
\end{equation}

\begin{figure}
\centering\includegraphics[ trim = {0 2cm 0 0}, width=0.9\linewidth]{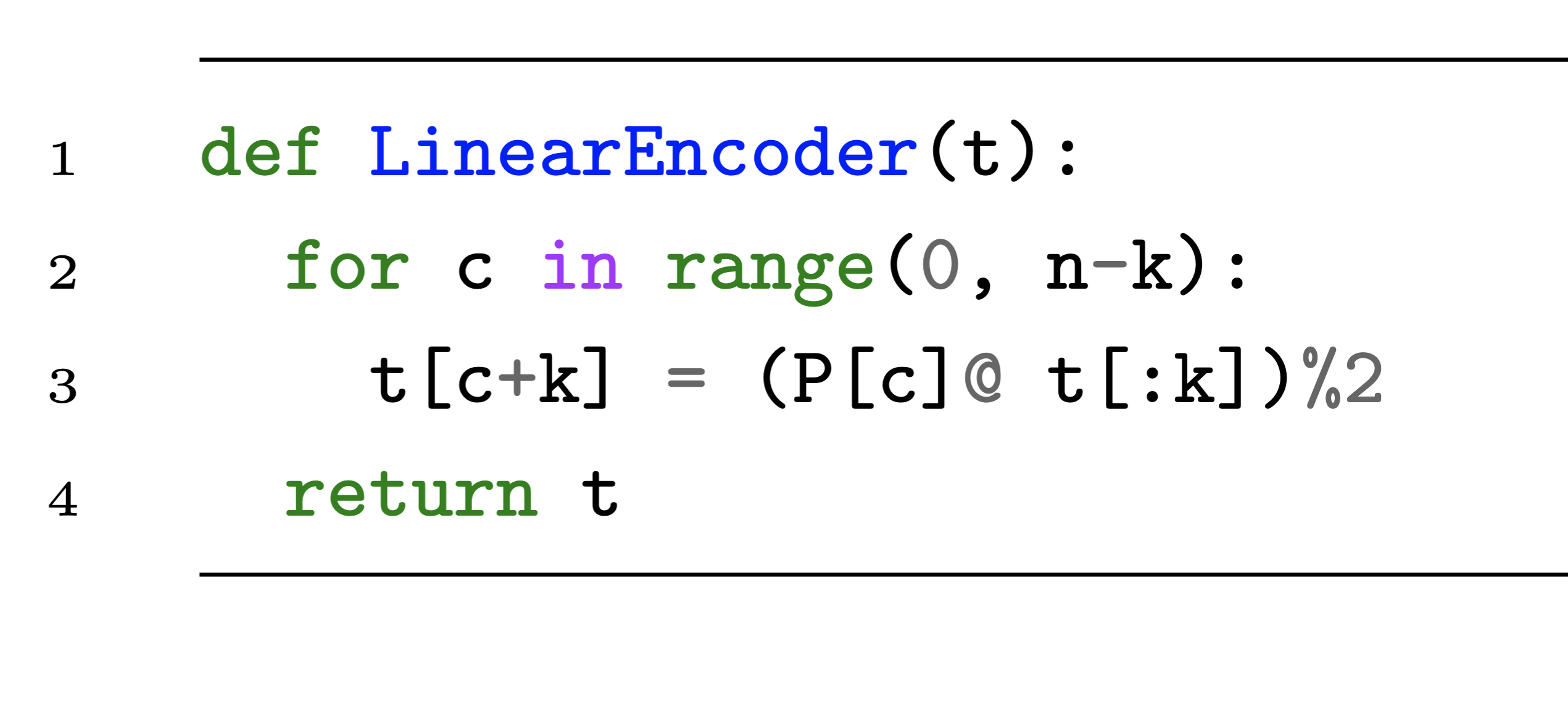}
\caption{Python script of a linear encoding program. The first $k$ bits of the vector $\mathbf{t}$ correspond to the input string, while the remaining bits are parity bits computed during program execution using the parity check matrix $P$. The variable $i$ serves as a loop counter, initialized to $i = 0$ at the start of the program.}\label{code:encoder}
\end{figure}. 

This transition of the distribution from $p_{X_r}$ to $p_{X_0}$, which occurs due to the reinitialization before the start of a new run, incurs a mismatch cost. The MMC associated with this reinitialization step is given by:
\begin{equation}
\MC^a = D(p_{X_r} \| q^a_{X_f}) - D(p_{X_0} \| \q^a_{X_f}),
\end{equation}
where $q^a_{X_f}$ is the prior distribution associated with the overwriting process. After the input and special variables are reinitialized, the prior evolves to:
\begin{equation}
\q^a_{X_f}(\x) = p_{X_{in}}(x_{in}) \delta_{X_{sp}}(x_{sp}) q^a_{X_{nin}}(x_{nin}).
\end{equation}
where $q^a_{X_{nin}}(x_{nin}) = \sum_{x_{in}, x_{sp}} q^a_{X_f}(\x)$ is the marginal distribution of $X_{nin}$.

Given this, the mismatch cost simplifies to:
\begin{equation}
\MC^a = D(p_{X_r} \| q^a_{X_r}) - D(p_{X_{nin}} \| q^a_{X_{nin}}).
\end{equation}
The above consideration of  mismatch cost of setting new values before reusing the computational machine extends to any physical device that involves overwriting. 

In the next couple of sections, we compute the mismatch cost associated with linear encoding and syndrome decoding algorithm. 

\subsection{Mismatch cost of linear encoding algorithm}\label{sec_MC_linear_codes}

Consider the linear encoding algorithm illustrated in Fig.~\ref{code:encoder}. It is an $(n, k)$ linear encoder with an associated parity-check matrix $P$,
\begin{equation}
    P = \begin{bmatrix}
        1 & 1 & 0 & 1 \\
        1 & 0 & 1 & 1 \\
        0 & 1 & 1 & 1 \\ 
        1 & 1 & 1 & 1
    \end{bmatrix}
\end{equation}
The algorithm takes as input an $n$-bit tape $\mathbf{t}$, where the first $k$ bits represent the input message $\mathbf{s}$, and the remaining $n-k$ bits are reserved for the parity bits. Inside the for-loop, the algorithm iteratively computes the $c^{th}$ parity bit and writes it to the $(c+k)^{th}$ position of $\mathbf{t}$. Values of $\mathbf{t}$ and the loop counter $c$ together constitute the state of the algorithm. 

\begin{figure*}
    \includegraphics[trim = {0 12cm 0 0}, width=1\textwidth]{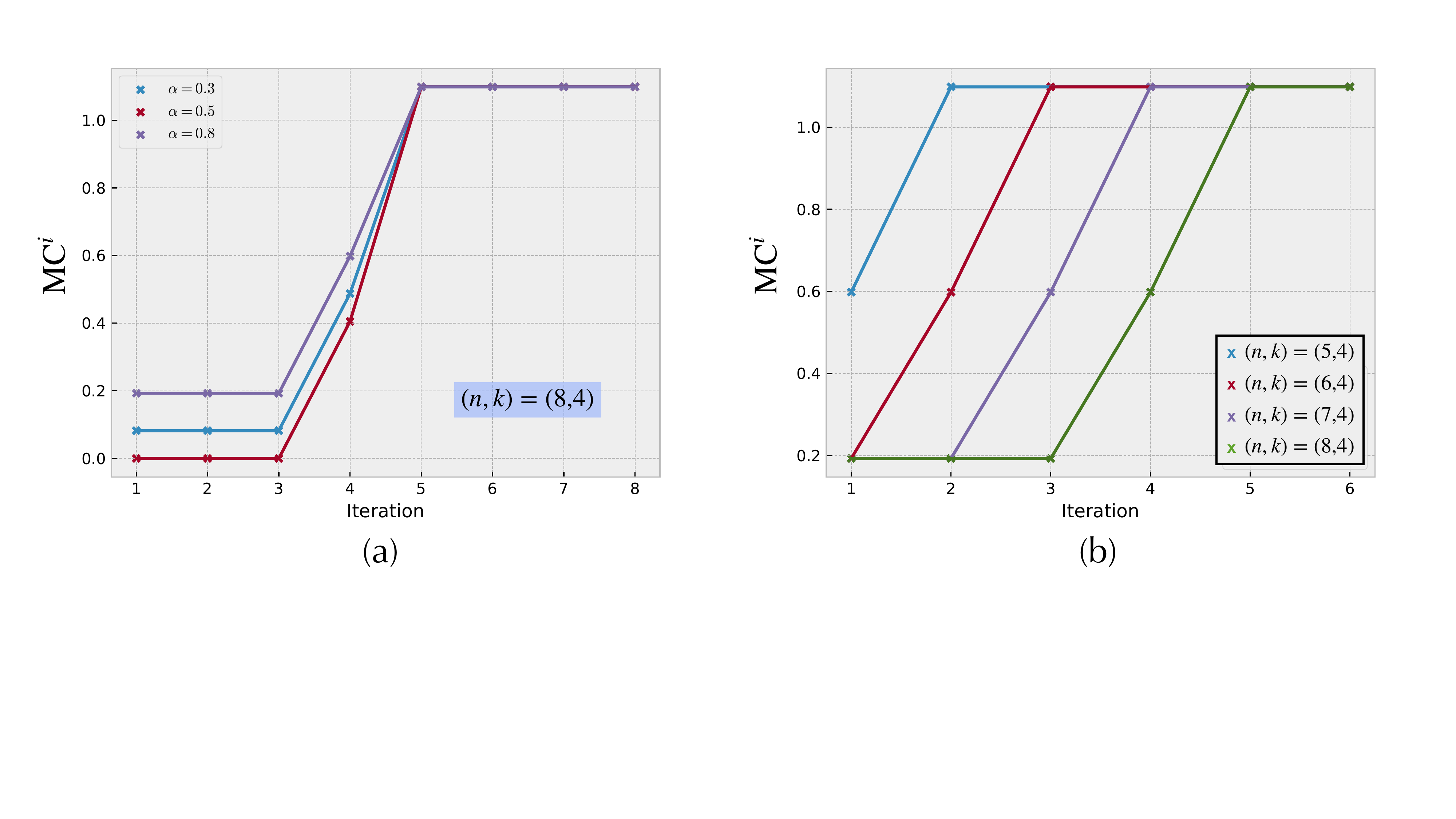}
\caption{Mismatch cost incurred in each iteration of the map $G_\E$ for (a) an (8,4) encoder. The input strings to the encoder are drawn from an i.i.d. Bernoulli distribution given in Eq.~\eqref{eq:input_dist}, with Bernoulli parameter $\alpha$. (b) Comparison of MMC incurred in each iteration across different encoders, where the encoding algorithms vary by the number of additional parity check bits. The input strings are sampled from a distribution with parameter \(\alpha = 0.80\) in (b).}
\label{fig:MC_encoder_iteration2}
\end{figure*}

To compute the MMC of this encoding algorithm, we construct the stochastic map  $G_\E$ associated with the algorithm, as described in Sec.~\ref{MC_of_Algorithms}. We assume that the $k$-bit input message $\mathbf{s}$ is sampled from an i.i.d. Bernoulli distribution with bias $\alpha$. That is,

\begin{equation}\label{eq:input_dist}
    p_{X_{in}}(\mathbf{s}) = \alpha^{\#_1(s)}(1-\alpha)^{\#_0(s)},
\end{equation}
where $\#_1(s)$ and $\#_0(s)$ denote the number of 1s and 0s in $\mathbf{s}$, respectively. The remaining $n-k$ bits of $\mathbf{t}$ retain their values from the previous execution of the algorithm, while the loop counter is re-initialized to $c = 0$. This defines the initial distribution $p_{X_0}$, as discussed in Section~\ref{sec:resetting_cost}.

After the $i^{th}$ step in the program, which corresponds to $i^{\text{th}}$ iteration of the for-loop which computes the $i^{\text{th}}$ parity bit, the distribution over the state of the algorithm is given by,

\begin{equation}
    p_{X_i} = G_{\E}^i p_{X_0}.
\end{equation}

To compute the mismatch cost at each step, we assume that the prior distribution associated with the physical process implementing $G_\E$ is uniform. Then, the mismatch cost incurred in the $i^{\text{th}}$ iteration is given by
\begin{equation}\label{eq: mck}
\MC^i (p_{X_0}) = D(G_\E^{i-1}p_{X_0}\|q_X) - D(G_\E^i p_{X_0}\|G_\E q_X).
\end{equation}

Completing the program requires $n - k$ iterations of the map $G_\E$. Therefore, the total mismatch cost (MMC) incurred by an $(n, k)$ linear encoding algorithm is given by

\begin{equation}\label{eq:62}
\MC_E(p_{X_0}) = \sum_{i = 1}^{n-k} \left[ D(G_\E^{i-1}p_{X_0} \,\|\, q_{X}) - D(G_\E^i p_{X_0} \,\|\, G_\E q_{X}) \right].
\end{equation}

Figure~\ref{fig:MC_encoder_iteration2}(a) shows the mismatch cost incurred in each iteration of a $(8, 4)$ linear encoding algorithm for various values of the Bernoulli parameter $\alpha$ of the input distribution. For a given $\alpha$, the mismatch cost remains constant across the first three iterations, which correspond to the computation of the first three parity bits. However, it increases during the computation of the fourth parity bit. In the fifth iteration, the algorithm halts, the system's dynamics reach a steady state, and the mismatch cost stabilizes.

Figure~\ref{fig:MC_encoder_iteration2}(b) compares the per-iteration MMC across different linear encoders, each with the same input length \(k = 4\) but varying numbers of appended parity bits. Interestingly, despite differences in total iterations and internal state space (e.g., tape length and loop counter size), certain patterns persist. For example, the MMC associated with computing the final parity bit is identical across encoders, whether it occurs in the first iteration of a $(5, 4)$ encoder or the fourth in a $(8, 4)$ encoder. Similar consistency appears for the second-to-last parity bit, third-to-last, and so on. Moreover, once the steady state is reached, all encoders exhibit the same constant MMC.

\begin{figure}
    \includegraphics[width= 1\linewidth]{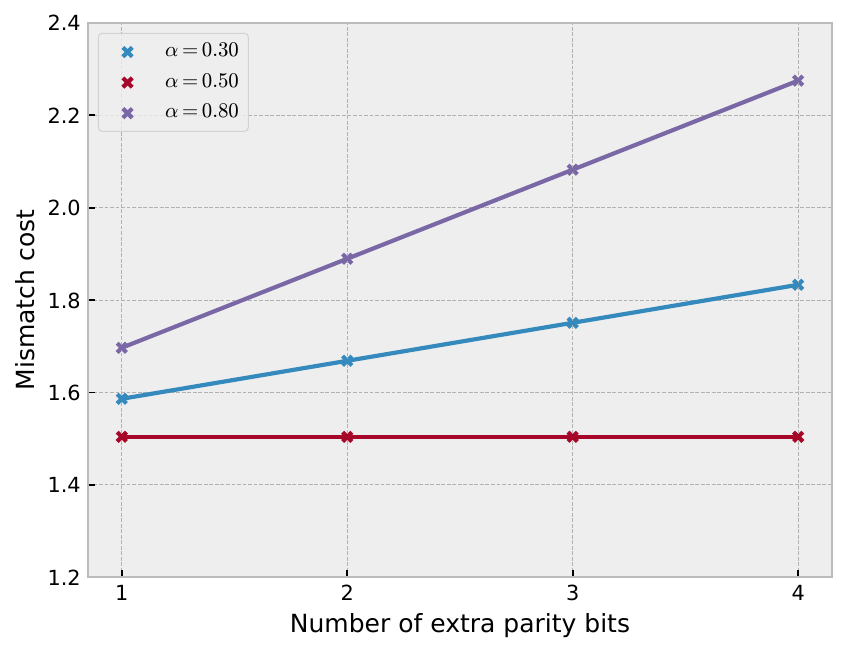}
    \caption{Total mismatch cost ($\MMC_E = \sum_{i}\MMC^i_E$) of the encoding algorithm for different encoding lengths. Here again, $k = 4$ is fixed and $n$ varies from $5$ to $8$. $\alpha$ is again the Bernoulli parameter of the probability distribution of the input strings~(see Eq.~\ref{eq:input_dist}.) }
    \label{fig:MC_encoder}
\end{figure}

Figure~\ref{fig:MC_encoder} shows the total mismatch cost, computed by summing over all iterations before the algorithm halts (as defined in Eq.~\ref{eq:62}), plotted against the number of parity bits for different values of the Bernoulli parameter \(\alpha\) of the input distribution. The results indicate that the total MMC grows linearly with the number of parity bits. This linear trend arises because each additional parity bit contributes a roughly constant mismatch cost, as observed in Figure~\ref{fig:MC_encoder_iteration2}(b). The magnitude of this cost depends on the input distribution: the further \(\alpha\) deviates from the uniform distribution, the higher the mismatch cost, as shown in Figure~\ref{fig:MC_encoder_iteration2}(a). 

In the next section, we perform a similar analysis for syndrome decoders. 

\begin{figure*}[ht]
    \includegraphics[trim = {0 1cm 0 0 }, width=0.8\textwidth]{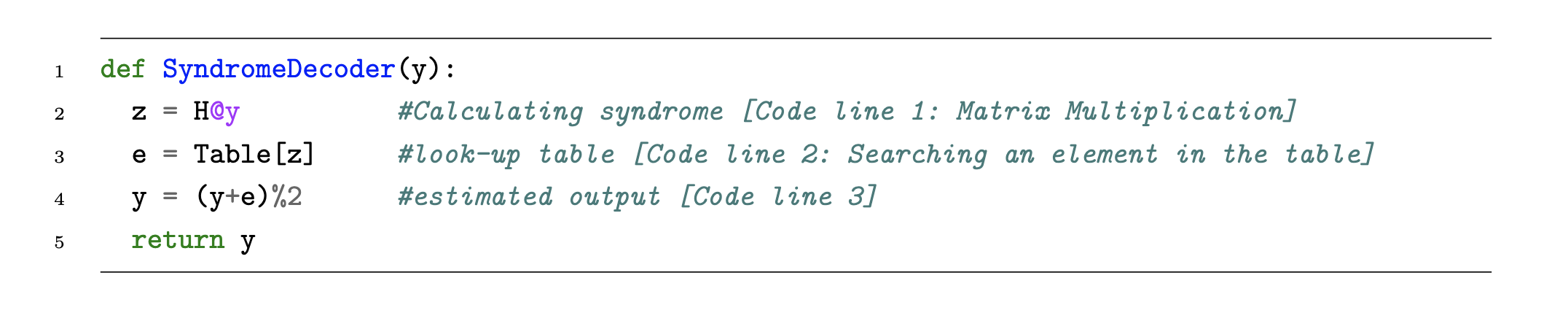}
    \caption{Python script for syndrome decoder. For a received $\mathbf{y}$, the first step involves computing its syndrome $\textbf{z}$ by using the parity check matrix $\Hh$. Based on the syndrome, error vector $\textbf{e}$ is estimated from pre-computed look up table in the second step. The third step involves correcting the received vector by adding the estimated error vector.}\label{code:decoder}
\end{figure*}

\begin{figure*}
    \includegraphics[trim = {0 12cm 0 0}, width=1\linewidth]{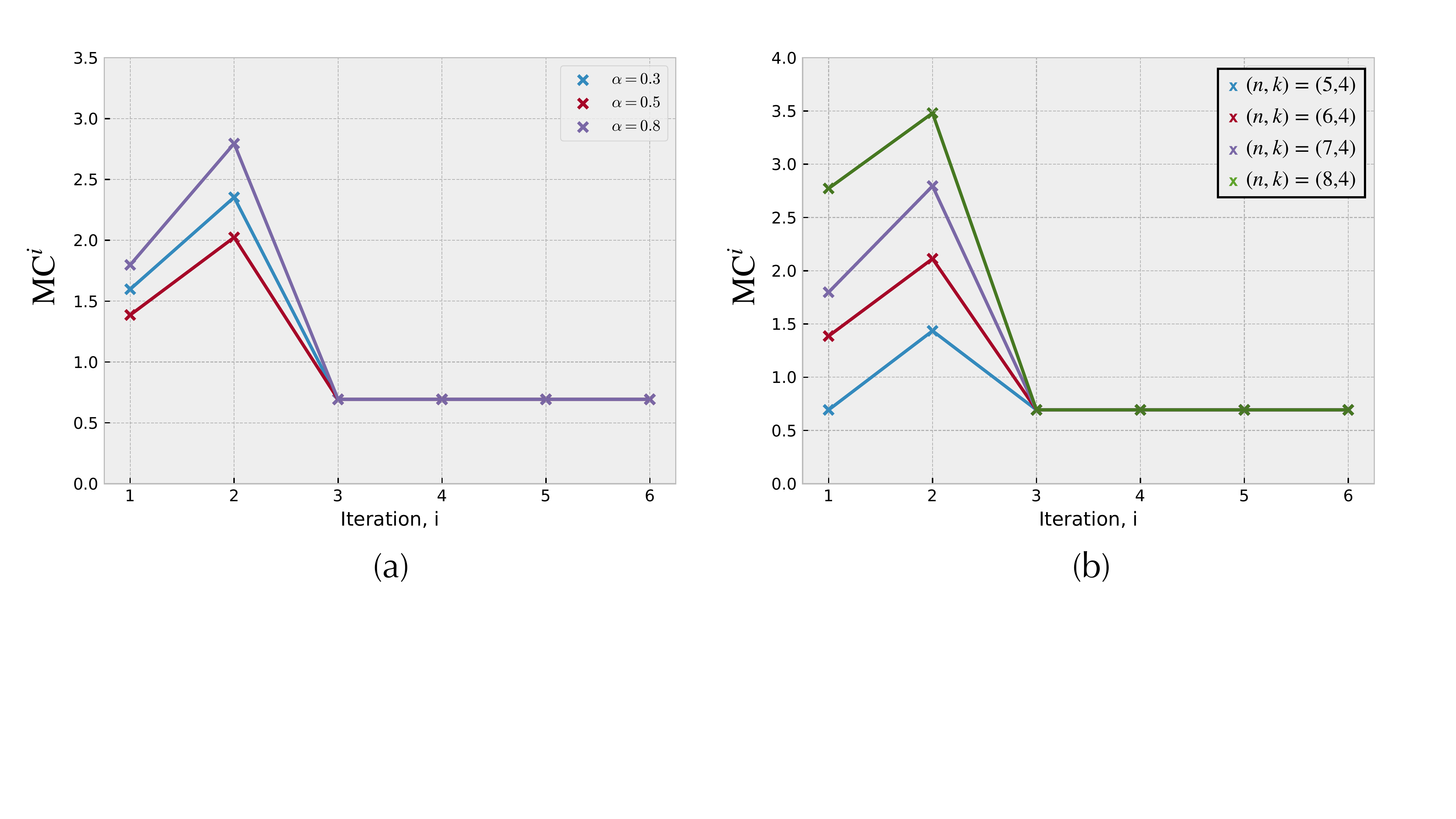}

\caption{(a) Mismatch cost in each step of a \((7, 4)\) syndrome decoder. The \( i^{th} \) iteration of \( G_D \) corresponds to the execution of the \( i^{th} \) step of the syndrome decoding algorithm (\ref{code:decoder}). The input string to the encoder is sampled from a distribution with Bernoulli parameter \(\alpha\). The encoded strings are transmitted through a binary symmetric noisy channel with noise level \( f = 0.001 \). The strings \(\mathbf{y}\) received at the decoder have a distribution that varies with changes in \(\alpha\). (b) Mismatch cost in each iteration of the map $G_D$ across decoders differing in the number of extra parity check bits $n-k$ used for decoding. Here, $k = 4$ and $\alpha = 0.80$.}
\label{fig:decoder2}
\end{figure*}

\subsection{Mismatch cost of syndrome decoding algorithm}
\label{sec_MC_syndrome_decoder}

We provide a high-level pseudo-code in Fig.~\ref{code:decoder} for a syndrome decoder, which operates on three main variables: $\mathbf{y}$, $\mathbf{z}$, and $\mathbf{e}$, along with a program counter denoted as $\pc$. Here, $\mathbf{y}$ is the $n$-bit string received from the channel, $\mathbf{z}$ is the computed syndrome, and $\mathbf{e}$ is the error vector. The program counter $\pc$ takes values 1, 2, and 3, corresponding to the three main steps of the decoding process: (1) computing the syndrome, (2) using a precomputed lookup table to identify the most probable error vector, and (3) correcting the received vector by adding the estimated error vector. 
Together, these variables define the state of the syndrome decoding algorithm, denoted as $(\mathbf{y}, \mathbf{z}, \mathbf{e}, \pc)$.   

The stochastic map $G_D$ associated with the decoding algorithm is constructed, as described in Sec.~\ref{MC_of_Algorithms}. We use $Y_i$ to denote the random variable representing the state of the decoder immediately after the $i^{th}$ iteration of the map. Additionally, $Y_0$ denotes the random variable representing the initial state of syndrome decoder, which is chosen as described in Sec.~\ref{sec:resetting_cost}: the program counter is set to 1 with probability 1; the non-input variables, $\mathbf{z}$ and $\mathbf{e}$, retain their values from the end of the previous execution; and the input variable $\mathbf{y}$ is freshly initialized by sampling from $ p_{Y_{\text{in}}}(\mathbf{y}) $, the distribution over received strings induced by the channel. $ p_{Y_{\text{in}}}(\mathbf{y}) $ is obtained from the input distribution to the encoder. Specifically, $p_{X_{\text{in}}}(\mathbf{s})$, defined in Eq.~\ref{eq:input_dist} for encoder, induces a distribution over the encoded codewords, which are transmitted through the noisy channel, thereby inducing a corresponding distribution over the received strings at the decoder input. Let $p_{Y_{\text{in}}}(\mathbf{y})$ denote the induced distribution over input to the decoder. 

To compute MMC associated with the decoder, we assume that the prior distribution $q_{X}$ for the decoding algorithm is uniform. MMC incurred in $i^{th}$ step of the decoding algorithm is given by, 
\begin{equation}
\MC^i_D(p_{Y_0}) = D(G_D^{i-1}p_{Y_0}|q_{Y})-D(G_D^ip_{Y_0}|G_D q_{Y})
\end{equation}

\begin{figure*}
    \includegraphics[trim = {0 12cm 0 0}, width=1\linewidth]{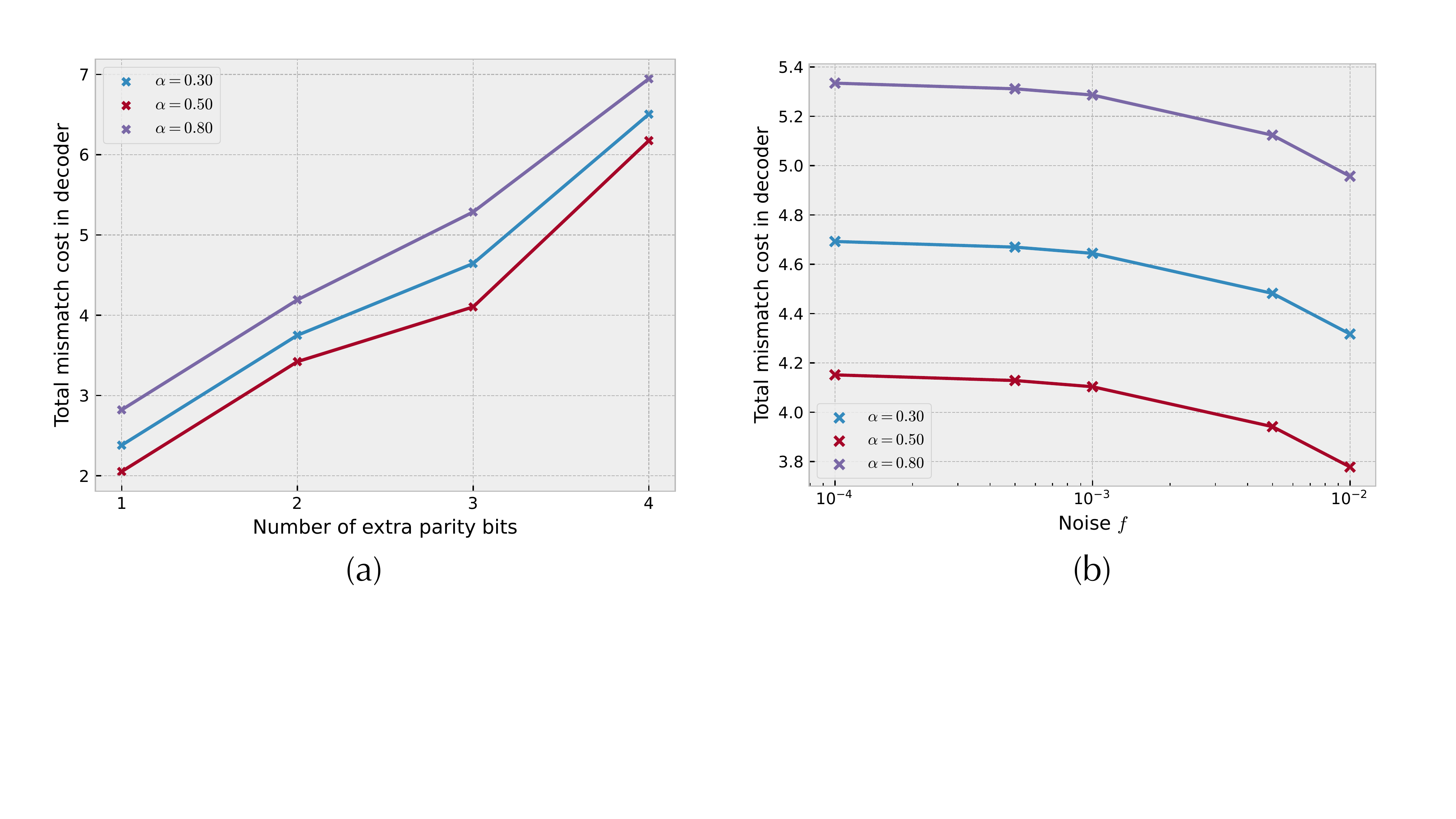}
    
\caption{(a) Total mismatch cost, $\MMC = \sum_{i = 1}^3 \MMC^i$, plotted against the number of parity bits $n-k$ in the codewords that the syndrome receives and decodes for different values of Bernoulli parameter $\alpha$. Here $k = 4$ and noise in the channel is symmetric with $\pi_{0|1} = \pi_{1|0} = 0.001$. (b) Total mismatch cost of a $(7, 4)$ syndrome decoder for various noise in the binary symmetric channel.}
\label{fig:decoder4}
\end{figure*}

\begin{figure*}
    \centering
    \includegraphics[trim = {0 25cm 0 0}, width=1.05\linewidth]{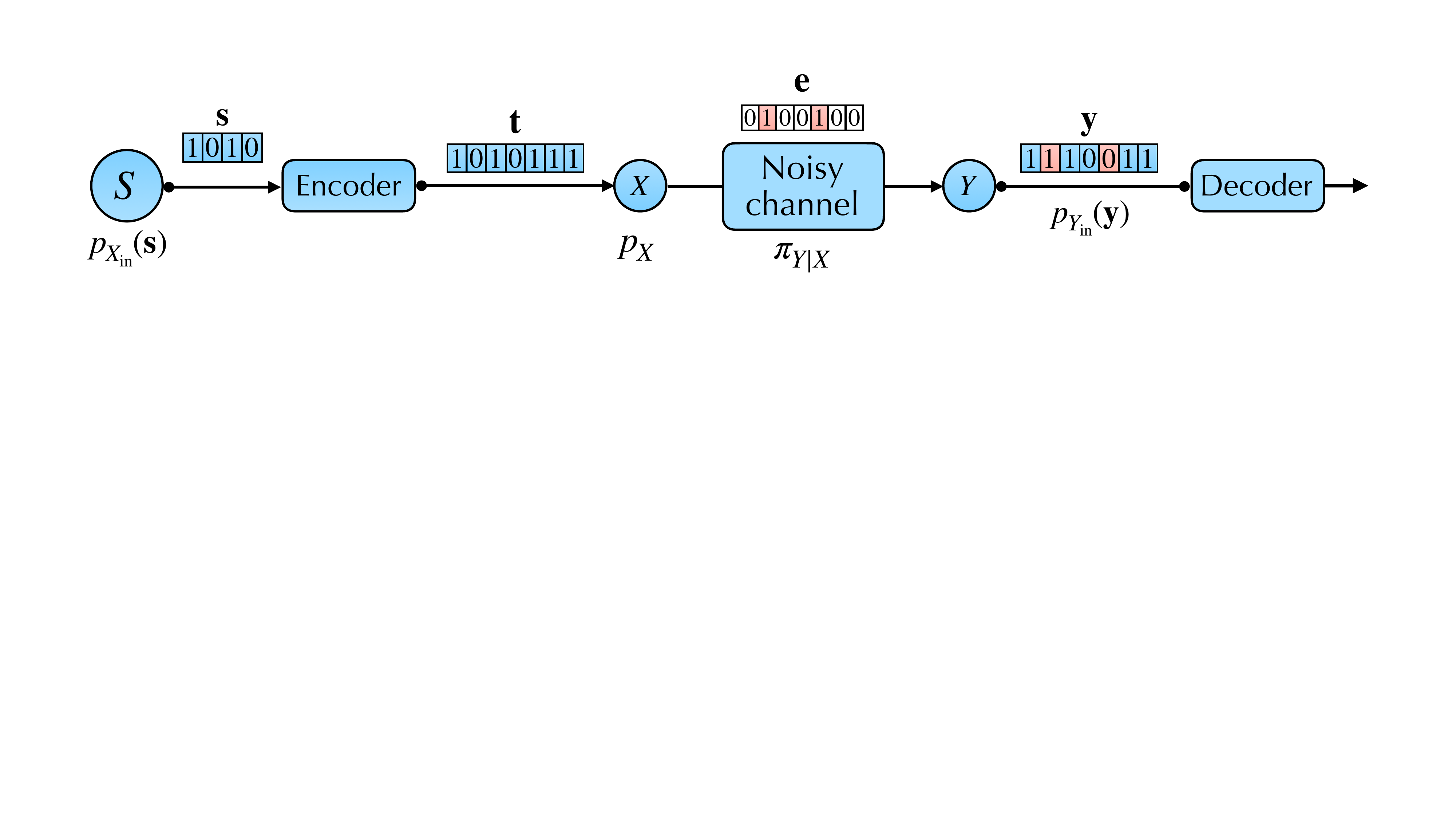}
        \caption{
    Full communication protocol: The source $S$ independently generates $k$-bit strings from a distribution $p_{X_{\text{in}}}$, which are encoded into $n$-bit codewords. Each encoded string is transmitted through a noisy channel one bit at a time, inducing a distribution $p_X$ at the channel input.}
    \label{fig:full_protocol}
\end{figure*}

Fig. \ref{fig:decoder2}(a) illustrates the mismatch cost in each iteration of a $(7, 4)$ syndrome decoder for various values of the Bernoulli parameter $\alpha$ of the input distribution. Since the decoding algorithm halts after three steps and the distribution $p_{Y_i}$ reaches a steady state, the mismatch cost stabilizes beyond the third iteration. In contrast to the encoding algorithm—where each step simply computes a parity bit—the computations performed during decoding differ across steps, resulting in significant variation in the mismatch cost. In particular, the second step, which updates $\mathbf{e}$ using a look-up table, incurs the highest mismatch cost.

Figure~\ref{fig:decoder2}(b) compares the per-iteration MMC across various $(n, 4)$ syndrome decoders, which differ in the value of $n$—that is, the length of the received $n$-bit string after a $k = 4$ bit input was encoded into an $n$-bit codeword and transmitted through a noisy channel.
All decoders reach a steady state after three iterations, independent of $n$. For larger $n$, the MMC in each decoding step is higher due to the greater number of parity bits involved. However, once the steady state is reached, the per-iteration MMC becomes identical across decoders.

\begin{figure*}[ht]
\includegraphics[trim = {0 16cm 0 0}, width=1\textwidth]{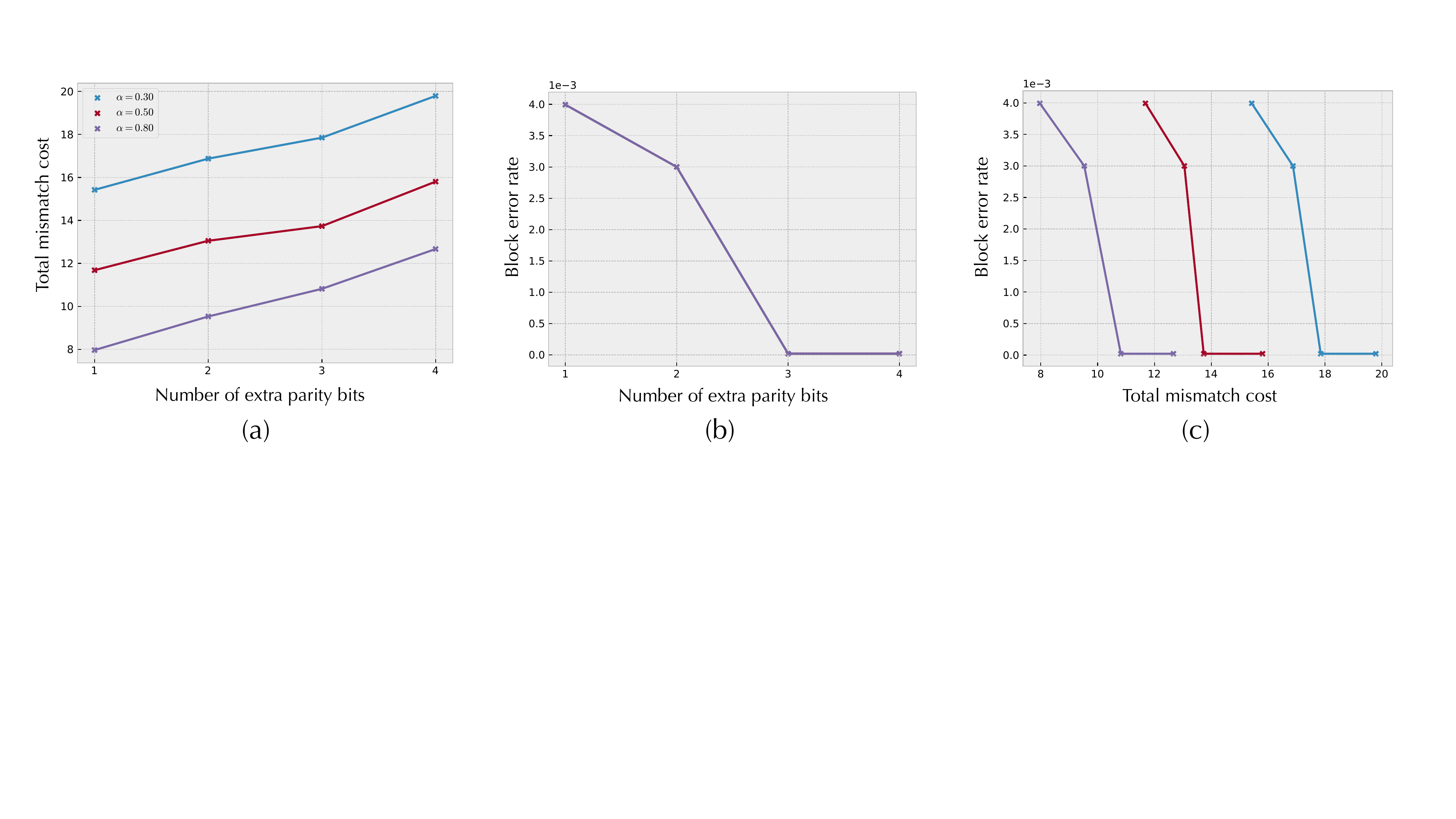}
\caption{(a) Total mismatch cost in an end-to-end communication protocol: $\MMC_E(p_{X_{in}})+\MMC_C(p_{X}) + \MMC_D(p_{Y_{in}})$. Here $k = 4$ and noise in the channel is symmetric with $\pi_{0|1} = \pi_{1|0} = 0.001$ and $\alpha$ is the Bernoulli parameter of input distribution $p_{X_{in}}$. (b) End-to-end block error rate decreases with the increasing number of parity check bits. (c) End-to-end block error rate vs. total mismatch cost. Lowering block error rate requires higher total MMC.}
 \label{fig:error_rate}
\end{figure*}

The total mismatch cost of syndrome decoder is obtained by summing over the three steps,

\begin{equation}\label{eq:67}
\MC_D(p_{Y_0}) = \sum_{k = 1}^{3} D(G_D^{k-1}p_{Y_0}\|q_{Y})-D(G_D^k p_{Y_0}\|G_D q_{Y}).
\end{equation}
Figure~\ref{fig:decoder4}(a) plots the total mismatch cost as a function of the number of parity bits $n - k$ in the codewords that the syndrome decoder receives and decodes. As expected, the total mismatch cost increases with the number of parity bits the decoder operates on. However, unlike the encoder case, this increase is not linear in $n$. 

We also examine how the mismatch cost varies with channel noise. Figure~\ref{fig:decoder4}(b) shows that the total mismatch cost of a $(7, 4)$ syndrome decoder decreases as the noise in a binary symmetric channel increases. This is because higher noise leads to a broader distribution over received strings $P_{Y_{in}}(\mathbf{y})$, making the induced distribution over the decoder's states closer to the assumed uniform prior. As a result, the mismatch cost, which quantifies deviation from the prior, decreases.

Finally, we combine the MMC of the encoder, the channel, and the decoder. The process begins with a source that i.i.d. generates messages from distribution $p_{X_{\text{in}}}$. These messages are passed through the encoder, producing encoded bit-strings that are transmitted through the channel one bit at a time. The encoded strings induce a distribution over the input node of the channel, denoted $p_X$. After transmission, the channel outputs distorted strings, which are received by the decoder and follow a distribution $p_{Y_{\text{in}}}$, which are then decoded. Three sequential three sequential components in the protocol—encoder, channel, and decoder—each repeatedly reused and incurring a mismatch cost: $\MMC_E(p_{X_{\text{in}}})$, $\MMC_C(p_X)$, and $\MMC_D(p_{Y_{\text{in}}})$, respectively.

The protocol operates cyclically: bit-strings generated by the source are passed through the encoder, transmitted through the channel, and then processed by the decoder. After each run, new bit-strings are generated, and the cycle repeats, with the inputs to the encoder, channel, and decoder being reinitialized sequentially. This repeated use introduces correlations between adjacent components (e.g., between the encoder output and the channel input) that are created during the forward pass and then broken during reinitialization. The resulting loss of mutual information between components in each cycle contributes to the total EP, over and above the sum of the mismatch costs of individual components. However, in this work, we do not analyze this inter-component contribution and focus solely on the sum of the mismatch costs from the encoder, channel, and decoder: $\MMC_E(p_{X_{in}})+\MMC_C(p_{X}) + \MMC_D(p_{Y_{in}})$. 

Figure~\ref{fig:error_rate}(a) shows how the total end-to-end mismatch cost increases with the number of parity bits used in the encoding and decoding process, for a fixed noisy channel and a given Bernoulli input distribution with parameter $\alpha$ (see Eq.~~\ref{eq:input_dist}). As expected, using more parity bits improves error correction, resulting in a lower end-to-end block error rate (Fig.~~\ref{fig:error_rate}(b)). Consequently, Fig.~\ref{fig:error_rate}(c) shows that block error rate decreases as total mismatch cost increases.

\section{Discussion and future directions}\label{Discussion}

In this paper, we have examined the MMC contribution to the thermodynamic cost of communication and, in doing so, established a minimal cost associated with communication channels. Our analysis centers on a general model of channel operation—namely, a repeated cycle of copy-reset-copy steps—which is both natural and ubiquitous across physical and engineered systems. Crucially, we showed that the minimal thermodynamic cost per communication cycle is lower bounded by the amount of information transmitted in that cycle. This result is universal: it holds independently of the specific physical implementation of the communication process.

Given the generality of the framework, we believe that the methods and results developed in this work can offer insight into a wide range of communication systems beyond engineered channels. In particular, they may help illuminate the fundamental thermodynamic costs of communication in biological systems—such as neural information processing in the brain, where neurons continuously exchange signals to process information~\cite{balasubramanian2015heterogeneity, balasubramanian2021brain}, or in cellular signaling networks that transmit biochemical messages~\cite{waltermann2011information, ten2016fundamental}, often at considerable thermodynamic cost. %In particular, in~\cite{ten2016fundamental} derives a mutual information lower bound on EP in 

In addition, our analysis may have implications for communication complexity theory and the study of distributed systems. Traditionally, communication complexity is quantified in terms of the number of messages exchanged or the number of communication rounds required to perform a distributed task~\cite{yao1979some, papadimitriou1982communication, newman1991private, yao1986generate}. We suggest that incorporating thermodynamic cost---as formalized through MMC---into this framework could enrich the field, potentially leading to new insights into the design of communication-efficient and energy-efficient distributed algorithms, as well as optimal network topologies. Additionally, we hope that our analysis will be relevant to the field of neuromorphic computing---an emerging computational architecture that, among other aims, seeks to overcome the substantial energetic cost associated with communication between memory and processor that traditional von Neumann architecture suffers from~\cite{burr2020emerging, markovic2020physics}.

With that said, we believe there is substantial room for future exploration and refinement. In this paper, we have exclusively focused on i.i.d. information sources, and the minimal thermodynamic costs we derive apply strictly under that assumption. However, many scenarios involve correlated signals, such as those generated by first- or second-order Markov processes. A natural question then arises: how does the minimal cost of communicating each bit change when the source exhibits memory or temporal correlations? Understanding this could yield a deeper theory of thermodynamic costs for correlated or memory-bearing sources.

Furthermore, as emphasized in our discussion of reverse-multiplexing, the relationship between mismatch cost and communication rate is not universally concave---it depends critically on the choice of prior distribution. This observation opens up two broad avenues for future work:
(i) For what classes of prior distributions does the mismatch cost exhibit concavity with respect to communication rate, and what constraints does this impose on the detail of the underlying process implementing that communication channel?
(ii) Beyond concavity, what other forms can this relationship take for different priors, and what novel communication strategies—analogous to, but distinct from, reverse-multiplexing—do these suggest?

We observed that the stepwise MMC incurred in both the linear encoder and the syndrome decoder increases with the length of the encoded string. However, this growth behaves differently for the two components: it is linear for the encoder, but nonlinear for the decoder.

The linear growth in the encoder's MMC reveals an important insight: as the encoding length increases, so does the number of computational steps—each parity bit adds a step. The fact that each additional step contributes a roughly constant increase in MMC, regardless of the expanding state space, suggests that the cost is primarily dictated by the nature of the computational step itself, which in this case is the computation of parity check bit, rather than the size of the underlying state space.

The decoding algorithm exhibits more complex behavior. As the encoding length grows, not only does the state space of the decoder grow, but so does the nature of its computation. For example, when calculating the syndrome using matrix multiplication, the dimensions of the matrices change with the length of the received string \(\mathbf{y}\), resulting in variations in the underlying computation. Consequently, the total MMC of syndrome decoder does not exhibit a linear growth with increasing length of encoded string. 

Additionally, while our numerical results consistently show that the second step of the decoding process dominates the total mismatch cost, we currently lack a clear analytical intuition for this behavior. This remains an interesting open question and points to a need for further investigation into the fine-grained structure of the decoder's thermodynamic costs.

As emphasized earlier, the stepwise MMC for the encoder and decoder presented in this work are derived under the assumption of a uniform prior distribution. This choice is intended purely for illustrative purposes; the results are not meant to imply universality, but rather to demonstrate that a detailed step-by-step analysis of minimal thermodynamic cost is feasible—even in the absence of an explicit physical model of the computational process.

Importantly, the periodic machine framework employed in this study can be readily applied to analyze the minimal costs associated with other encoding and decoding protocols, such as turbo codes or LDPC codes. However, a significant open question remains: Are there universal features of minimal thermodynamic costs for encoding and decoding that hold irrespective of the prior distribution, depending only on the structure or complexity of the protocol itself?

A potential direction for discovering prior-independent minimal costs involves analyzing the structure of the periodic mismatch cost itself. Consider Eq.~\ref{eq_sum_MMC_periodic}, which defines the total mismatch cost for a given prior $q_X$:

\begin{equation}
 \MC_{q_X}(p_{X_0}) = \sum_{i = 1}^{r-1} \left[D(G^{i-1}p_{X_0} \| q_X) - D(G^i p_{X_0} \| Gq_X)\right]
\end{equation}

As the system evolves through a sequence of state distributions $\{p_{X_0}, p_{X_1}, \dots, p_{X_r}\}$ under the update rule $p_{X_{t+1}} = G p_{X_t}$, one can find a distribution $\hat{q}_X$ that minimizes of the sum on the right-hand side, and yields a special MMC value, denoted $\MMC_{\hat{q}_X}(p_{X_0})$, that serves as a strictly positive lower bound on the MMC incurred for any other choice of prior $q_X$. 

While this observation suggests a promising direction for deriving prior-independent minimal thermodynamic cost, we do not pursue it further in this paper and leave a detailed development for future work.

\section{Acknowledgement}

The code used in this study is openly available at~\cite{yadav2025github}.
This work was supported by the U.S. National Science Foundation (NSF) Grant 2221345. We would like to thank Francesco Caravelli for helpful discussions. We would also like to thank the Santa Fe Institute for helping to support this research.

\appendix
\onecolumngrid

\section{Derivations}
\subsection{Subsystem Processes}\label{App1}

Consider a system composed of two subsystems $A$ and $B$ with state space $\X$ and $\Y$ respectively. A process is called a \textit{ subsystem process}~\cite{wolpert2020uncertainty, wolpert2019stochastic} if for any initial distribution $p_{XY}(x, y)$ over the joint state space, the following two conditions are satisfied:
\begin{itemize}
    \item[(i)] Two subsystems evolve independent of each other, i.e.,
    \begin{equation}\label{eq: sub1}
        G_{AB}(x_1, y_1 | x_0, y_0) = G_A(x_1 | x_0)G_B(y_1 | y_0),
    \end{equation}
    \item[(ii)] Total entropy flow of the joint system, denoted as $Q_{AB}$ is the sum of the entropy flow of the subsystems,
    \begin{equation}\label{eq: sub2}
        Q_{AB}(p_{XY}) = Q_A(p_X)+Q_B(p_Y).
    \end{equation}
\end{itemize}
According to equation (\ref{EP_def}), the total EP of the joint system is given by,
\begin{equation}\label{eq:subsystem_EP1}
    \C(p_{XY}) = Q(p_{XY}) + S(G_{AB}p_{XY}) - S(p_{XY}).
\end{equation}
On the other hand, EP of individual subsystems can be written as,
\begin{equation}
    \C_A(p_X) = Q^A(p_0^A) + S(G^{A}p^{A}_0) - S(p^{A}_0).
\end{equation}
and likewise for $\C_B$. Therefore, we can write (\ref{eq:subsystem_EP1}) as
\begin{align}\label{eq:subsystem_EP2}
     \C &= \C_A + \C_B + \left( S(p_{X}) + S(p_{Y})-S(p_{XY})\right) - 
     \left(S(G_{A}p_{X}) + S(G_{B}p_{Y}) -S(G_{AB}p_{XY})\right) \\
    &= \C_A + \C_B + I_0(X;Y) - I_1(X;Y) \\
      &= \C_A + \C_B + \Delta \I(X; Y) \label{eq:15}
\end{align}
where $\Delta \I(X; Y) = \I_0(X; Y) -\I_1(X; Y)$ denotes the drop in mutual information between $X$ and $Y$ from the initial to the final distribution. By the data processing inequality, mutual information cannot increase under local processing: $\I_0(X; Y) \ge \I_1(X; Y)$. Consequently, in a subsystem process, even if each subsystem evolves thermodynamically reversibly—i.e., with zero entropy production in each part ($\C_X = \C_Y = 0$)—the joint system may still exhibit a positive total entropy production due to a loss in mutual information between the subsystems. Specifically, if the initial correlation between subsystems $X$ and $Y$ is lost during the process, then the drop in mutual information contributes positively to total EP:
\begin{equation}
\C \ge \Delta \I(X : Y).
\end{equation}
However, if the initial joint distribution is uncorrelated—i.e., a product distribution—then $\I_0(X; Y) = 0$, and hence $\Delta \I(X; Y) = 0$. In this case, if $q_X$ and $q_Y$ are the prior distributions minimizing the entropy production in the $X$ and $Y$ subsystems respectively, then the joint prior minimizing the total EP is simply their product:
\begin{equation}
q_{XY}(x, y) = q_X(x) q_Y(y).
\end{equation}
\qed
% \begin{enumerate}
%     \item \begin{equation}
%         \sum_{x}q_{XY}(x, y) = \mathrm{arg}\min_{p_X}\C_X(p_X) \mathrm{and} \sum_{y}q_{XY}(x, y) = \mathrm{arg}\min_{p_Y}\C_Y(p_Y) 
%     \end{equation}
%     \item \begin{equation}
%         q_{XY}(x,y) = q_X(x)q_Y(y)
%     \end{equation}
% \end{enumerate}

% is a product distribution which guarantees that $\I_0(X;Y) = 0$. 

% \noindent
% Eq. (\ref{eq:subsystem_EP2}), which applies to two subsystems, can be extended to systems composed of multiple subsystems~\cite{wolpert2020uncertainty}. 

\subsection{Derivation of Eq.~\ref{eq:owMC} and Eq.~\ref{eq:52}}\label{App2}
During the input overwriting process, the actual joint distribution evolves from the correlated form $p^b_{XY}(x, y) = \pi_{Y|X}(y|x)p_X(x)$ to the product form $p^a_{XY}(x, y) = p_X(x)p_Y(y)$:
\begin{equation}
\pi_{Y|X}(y|x)p_X(x) \xrightarrow{(a)} p_X(x)p_Y(y).
\end{equation}
In parallel, the prior distribution evolves from $q^a_{XY}(x, y) = q^a_X(x)q^a_Y(y)$ to $\q^a_{XY}(x, y) = p_X(x)q^a_Y(y)$:
\begin{equation}
q^a_X(x)q^a_Y(y) \xrightarrow{(a)} p_X(x)q^a_Y(y).
\end{equation}
Thus, the corresponding drop in KL divergence during the process is given by:
\begin{align}
\MC^{a}(p_X) &= D\left(p^b_{XY}\| q^a_{XY}\right)- D\left(p^a_{XY}\| \q^a_{XY}\right) \\
    &= \sum_{x, y} \pi_{Y|X}(y|x) p_X(x) \ln \frac{\pi_{Y|X}(y|x) p_X(x)}{q^a_X(x)q^a_Y(y)}  - \sum_{x,y} p_X(x)p_Y(y)  \ln \frac{p_X(x)p_Y(y)}{p_X(x)q^a_Y(y)} \\
    &= \sum_{x, y} \pi_{Y|X}(y|x) p_X(x) \ln \frac{\pi_{Y|X}(y|x) p_X(x)}{p_X(x)p_Y(y)} + \sum_{x, y} \pi_{Y|X}(y|x) p_X(x) \ln \frac{p_X(x)p_Y(y)}{q^a_X(x)q^a_Y(y)}  \\ \nonumber
    &\qquad- \sum_{x,y} p_X(x)p_Y(y)  \ln \frac{p_Y(y)}{q^a_Y(y)}
\end{align}
In the third equality, the first term corresponds to the mutual information between $X$ and $Y$, while the second term decomposes into the sum of two KL-divergences: one between $p_X$ and $q^a_X$, and the other between $p_Y$ and $q^a_Y$. The third term is itself a KL divergence between $p_Y$ and $q^a_Y$. Therefore, the contribution involving $p_Y$ and $q^a_Y$ cancels out, leaving,
\begin{align}
\MC^{a}(p_X)
    &= \I(X;Y) + D(p_X\|q^a_X) + D(p_Y\|q^a_Y) -    D(p_Y\|q^a_Y) \\
    &= \I(X;Y) + D\left(p_X\|q^a_X\right)
\end{align}
\qed

Now, during the noisy copy operation, the actual distribution transforms from $p^a_{XY}(x,y) = p_X(x)p_Y(y)$ to $p^b_{XY}(x,y) = \pi_{Y|X}(y|x)p_X(x)$,
\begin{equation}
     p_X(x)p_Y(y) \xrightarrow{(b)} \pi_{Y|X}(y|x)p_X(x).
\end{equation}
while the prior distribution evolves from $q^b_{XY}(x, y)$ to $\q^b_{XY}(x,y) = \pi_{Y|X}(y|x)q^b_X(x)$,
\begin{equation}
    q^b_{XY}(x, y) \xrightarrow{(b)} \pi_{Y|X}(y|x)q^b_X(x).
\end{equation}
Therefore, the corresponding drop in KL-divergence during the copying process is given by:
\begin{align}
    \MMC^b(p_X) &= D(p^a_{XY}\| q^b_{XY}) - D(p^b_{XY}\|\q^b_{XY}) \\ 
    &= \sum_{x,y} p_X(x)p_Y(y)  \ln \frac{p_X(x)p_Y(y)}{q^b_{XY}(x, y)} - \sum_{x, y} \pi_{Y|X}(y|x) p_X(x) \ln \frac{\pi_{Y|X}(y|x) p_X(x)}{\pi_{Y|X}(y|x) q^b_X(x)}  \\
    &= D(p_{X}p_{Y}\| q^b_{XY}) - D(p_X\|q^b_X)
\end{align}
\qed
\subsection{Additivity MMC for multiple channels}\label{App3}

We focus on the case of two channels but the analysis below can be generalized to more than two channels. Consider two channels with input nodes $X_1$ and $X_2$, and output nodes $Y_1$ and $Y_2$, and each with condition distributions $\pi_{Y_1|X_1}$ and $\pi_{Y_1|X_1}$. We assume that for the process of overwriting of new input values, the input nodes $X_1$ and $X_2$ are independently sampled from $p_{X_1}$ and $p_{X_2}$ respectively. We aim to show that in this case, the MMC associated with the communication across the joint channel $\left((X_1, X_2), (Y_1, Y_2)\right)$ is the sum of MMC associated with the communication across the individual channels. We begin with the MMC of overwriting process. From Eq.~\ref{eq:owMC}, we have
\begin{equation}
    \MMC^a(p_{X_1 X_2}) = \I(X_1, X_2;Y_1, Y_2) + D\left(p_{X_1 X_2}\|q^a_{X_1 X_2}\right).
\end{equation}
where, since inputs are sampled independently from $p_{X_1}$ and $p_{X_2}$, the joint input distribution is a product distribution, $p_{X_1 X_2}(x_1, x_2) = p_{X_1}(x_1)p_{X_2}(x_2)$. This allows us to write down,
\begin{equation}\label{eq_App31}
    \I(X_1, X_2; Y_1, Y_2) = \I(X_1; Y_1) + \I(X_2; Y_2).
\end{equation}
Now, since the two input nodes $X_1$ and $X_2$ evolve independently during the overwriting process, as we discussed in~\ref{App1}, the prior distribution $q^a_{X_1 X_2}$ is a product distribution of the form 
$q^a_{X_1 X_2}(x_1, x_2) = q^a_{X_1}(x_1)q^a_{X_2}(x_2)$. Therefore, the KL-divergence term decomposes into two KL-divergences: 
\begin{equation}\label{eq_App32}
    D\left(p_{X_1 X_2}\|q^a_{X_1 X_2}\right)  = D\left(p_{X_1}\|q^a_{X_1}\right) + D\left(p_{X_2}\|q^a_{X_2}\right).
\end{equation}
Using Eq.~\ref{eq_App31} and Eq.~\ref{eq_App32}, we get a decomposition of MMC of joint overwriting process into sum of MMC of overwriting processes for each input node,
\begin{equation}\label{eq_App33}
    \MMC^a(p_{X_1 X_2}) = \MMC^a(p_{X_1}) + \MMC^a(p_{X_2})
\end{equation}

Now, according to Eq.~\ref{eq:52}, MMC of copy operation is given by,
\begin{equation}\label{eq_App34}
    \MMC^b(p_{X_1 X_2}) = D(p_{X_1 X_2}p_{Y_1 Y_2}\| q^b_{X_1 X_2 Y_1 Y_2}) - D(p_{X_1 X_2}\|q^b_{X_1 X_2}).
\end{equation}
Since the output values update according to conditional distributions $\pi_{Y_1|X_1}$ and $\pi_{Y_2|X_2}$ separately, therefore,
\begin{align}
    p_{Y_1 Y_2}(y_1, y_2) &= \sum_{x_1, x_2} \pi_{Y_1|X_1}(y_1|x_1) \pi_{Y_2|X_2}(y_2|x_2) p_{x_1}(x_1)p_{x_2}(x_2) \\
    &= \left(\sum_{x_1} \pi_{Y_1|X_1}(y_1|x_1)p_{X_1}(x_1)\right) \left(\sum_{x_2} \pi_{Y_2|X_2}(y_2|x_2)p_{X_2}(x_2)\right) \\
    &= p_{Y_1}(y_1) p_{Y_2}(y_2) \label{eq_App35}
\end{align}
Now, since the two joint systems $(X_1, Y_1)$ and $(X_2, Y_2)$ evolve independently during the copy operation according to the conditional distribution $\pi_{Y_1|X_1}$ and $\pi_{Y_2|X_2}$, therefore, the prior distribution $q^b_{X_1 X_2 Y_1 Y_2}$ decomposes into product distributions: 
\begin{equation}\label{eq_App36}
    q^b_{X_1 X_2 Y_1 Y_2}(x_1, x_2, y_1, y_2) = q^b_{X_1 Y_1}(x_1, y_1)q^b_{X_2 Y_2}(x_2, y_2).
\end{equation}
Using Eq.~\ref{eq_App35} and~\ref{eq_App36}, the first KL-divergence term in~\ref{eq_App34} decomposes into two KL-divergences:
\begin{equation}\label{eq_App37}
    D(p_{X_1 X_2}p_{Y_1 Y_2}\| q^b_{X_1 X_2 Y_1 Y_2}) = D(p_{X_1}p_{Y_1}\| q^b_{X_1 Y_1}) + D(p_{X_2}p_{Y_2}\| q^b_{X_2 Y_2})
\end{equation}
Now, $q^b_{X_1 X_2}(x_1, x_2) = \sum_{y_1, y_2}q^b_{X_1 Y_1}(x_1, y_1)q^b_{X_2 Y_2}(x_2, y_2) = q^b_{X_1}(x_1)q^b_{X_2}(x_2)$, and therefore the second KL-divergence in~\ref{eq_App34} also decomposes into sum of two terms,
\begin{equation}\label{eq_App38}
    D(p_{X_1 X_2}\|q^b_{X_1 X_2}) = D(p_{X_1}\|q^b_{X_1}) + D(p_{X_2}\|q^b_{X_2}).
\end{equation}
Eq.~\ref{eq_App37} and~\ref{eq_App38} allow us to write the MMC of the joint copy operation as a sum of MMC of individual copy operations: 
\begin{equation}\label{eq_App39}
    \MMC^b(p_{X_1 X_2}) = \MMC^b(p_{X_1}) + \MMC^b(p_{X_2}).
\end{equation}
Hence, Eq.~\ref{eq_App39} and~\ref{eq_App33} tell us that when the values of the inputs of the two channels are drawn independently and when the outputs each channel evolves according to $\pi_{Y_1|X_1}$ and $\pi_{Y_2|X_2}$, then the MMC of communication across the joint channel is the sum of the MMC of communication across each channel considered separately:
\begin{equation}
    \MMC(p_{X_1}p_{X_2}) = \MMC(p_{X_1})+ \MMC(p_{X_2})
\end{equation}
\qed

%\subsection{MMC of channel in which output resets before new input is copied.}\label{App14}
\end{document}